\documentclass[12pt]{article}

\usepackage[margin=1in]{geometry}
\usepackage{setspace}
\usepackage{enumitem}
\usepackage{mathtools}
\usepackage{amsthm}
\usepackage{dsfont}
\usepackage{amssymb}
\usepackage{dsfont}
\usepackage{bm}
\usepackage{breqn}
\usepackage{color}
\usepackage{tikz}
\usetikzlibrary{patterns,decorations.pathreplacing}
\usepackage[authoryear]{natbib}
\usepackage[hidelinks,breaklinks]{hyperref}
\usepackage[flushleft]{threeparttable}

\allowdisplaybreaks

\newtheorem{proposition}{Proposition}

\newtheorem{reparameterization}{Reparameterization}
\newtheorem{definition}{Definition}

\newtheoremstyle{named}{}{}{\itshape}{}{\bfseries}{.}{.5em}{#1\thmnote{ #3}}
\theoremstyle{named}

\newcommand{\inv}{^{-1}}
\newcommand{\isum}{\sum_{i=1}^n}

\newcommand{\R}{\mathbb{R}}

\newcommand{\shrinkageparameter}{1}

\begin{document}

\title{Weak Identification in Low-Dimensional Factor Models with One or Two Factors\footnote{This research has benefited from comments from many conferences, seminars, and occasional readers. I want to especially acknowledge helpful feedback/advising from Donald Andrews, Jushan Bai, Xiaohong Chen, Yuichi Kitamura, and Serena Ng.}
}
\author{Gregory Fletcher Cox\\Department of Economics, National University of Singapore
}
\date{
\today}

\maketitle

\begin{abstract}
\normalsize 
This paper describes how to reparameterize low-dimensional factor models with one or two factors to fit weak identification theory developed for generalized method of moments models. 
Some identification-robust tests, here called ``plug-in'' tests, require a reparameterization to distinguish weakly identified parameters from strongly identified parameters. 
The reparameterizations in this paper make plug-in tests available for subvector hypotheses in low-dimensional factor models with one or two factors. 
Simulations show that the plug-in tests are less conservative than identification-robust tests that use the original parameterization. 
An empirical application to a factor model of parental investments in children is included. \vspace{-0.1cm}
\end{abstract}



\pagebreak
\section{Introduction}

Low-dimensional factor models are commonly used in many fields to account for latent variables.\footnote{Textbook treatments of low-dimensional factor models, such as \cite{Harman1976}, \cite{Thompson2004}, \cite{Gorsuch2014}, and \cite{Brown2015}, include examples from economics, political science, sociology, psychology, and biomedical sciences.} 
In economics, low-dimensional factor models are commonly used to account for unobserved skills or abilities in the literature following \cite{CarneiroHansenHeckman2003}, \cite{HeckmanStixrudUrzua2006}, \cite{CunhaHeckman2008}, \cite{CunhaHeckmanSchennach2010}, and \cite{HeckmanPintoSavelyev2013}. 
Conditions for identification in low-dimensional factor models have been derived in \cite{AndersonRubin1956}. 
These conditions indicate problematic points in the parameter space where identification is lost. 
Similar to an instrumental variables model with a weak instrument, weak identification arises when the true value of the parameters is close, in some sense, to one of these problematic points. 
As with weak instruments, weak identification implies that standard hypothesis tests, including t-tests and Wald tests, are invalid. 
Several papers address the weak identification problem in generalized methods of moments (GMM) models, including \cite{StockWright2000} and \cite{Kleibergen2005}. 
This paper describes how to reparameterize low-dimensional factor models to fit weak identification theory developed for GMM models. 

There are two types of identification-robust tests for subvector hypotheses in the weak identification literature: plug-in tests that require strongly identified nuisance parameters, such as the K test from \cite{Kleibergen2005}, and other tests that allow for weakly identified nuisance parameters, such as the test in \cite{ChaudhuriZivot2011}. 
A reparameterization that classifies each parameter as weakly identified or strongly identified makes the plug-in tests available for identification-robust inference. 
Otherwise, those tests would have to be projected over the nuisance parameters, making them very conservative. 

Reparameterizations that satisfy the weak identification classification can be challenging to find. 
Most of the nonlinear models for which this classification has been solved have relatively few parameters. 
For example, \cite{AndrewsMikusheva2016Geometric} verify the weak identification classification in a simplified small-scale DSGE model with six parameters. 
\cite{AndrewsMikusheva2016Geometric} state, ``even in this simple highly stylized model, deriving the weakly and strongly identified directions in the parameter space is messy, and such derivations will be difficult if not impossible in richer, more empirically relevant models.''\footnote{\cite{AndrewsMikusheva2016Geometric}, Section S8, pg. 27.} 
In this paper, the low-dimensional factor models that we classify constitute a class of empirically relevant models. 

This paper focuses on weak identification in low-dimensional factor models with one or two factors. 
In the two-factor model, we also focus on the case that one of the factors has at least three nonzero factor loadings. 
Section A.1, in the Supplemental Materials, discusses the possibility of extending the results in this paper to the case of three factors or the case of two factors with neither factor assumed to have three nonzero factor loadings. 

We include simulations comparing identification-robust hypothesis tests. 
We find that the plug-in tests have good size and power properties. 
In contrast, tests that allow for weakly identified nuisance parameters can be very conservative under weak identification. 
We also document the fact that the number of factors is identified under weaker conditions than the model parameters. 
Estimates of the number of factors may estimate the correct number of factors even if the factor model is weakly identified. 
These simulations complement simulations by \cite{BriggsMacCallum2003} and \cite{Ximenez2006}, which compare estimators of the factors when the factor loadings are near zero. 

Empirical applications of factor models are often weakly identified. 
Anytime the number of factors is unknown, the researcher must consider the possibility that the factor model is weakly identified. 
\cite{Attanasio2020AER} use a factor model with one factor to model parental investments in children. 
We note evidence of a second factor, indicating that their model is misspecified. 
In a two factor model, we compute identification-robust confidence intervals. 

\cite{HanMcCloskey2019} provide a strategy for finding a reparameterization based on solving a sequence of differential equations. 
Assumption ID in \cite{HanMcCloskey2019} requires a subset of the parameters to be strongly identified. 
Then, \cite{HanMcCloskey2019} recommend solving a sequence of differential equations to classify the remaining parameters. 
The low-dimensional factor models considered in this paper do not satisfy Assumption ID in their original parameterization; see Remark 3.3, below. 

\cite{Cox_weak_id_w_bounds} uses these reparameterizations to analyze the combination of weak identification with bounds in low-dimensional factor models. 
The idea is that inequalities on the parameters can shrink the identified set and confidence intervals. 
\cite{Cox_weak_id_w_bounds} proposes an identification-robust quasi-likelihood ratio test that uses information from the inequalities when identification is weak. 

It should be pointed out that weak identification in low-dimensional factor models is different from weak factors in high-dimensional factor models, as in \cite{Onatski2012} or \cite{Freyaldenhoven2022}. 
In high-dimensional factor models, the primary problem is an accumulation of noise from an increasing number of variables that do not have much information about the factors. 
In low-dimensional factor models, the problem is loss of identification. 
In fact, any factor in a low-dimensional factor model, with a bounded number of nonzero factor loadings, would count as a weak factor in a high-dimensional factor model.\footnote{This comparison holds the information about the factor equal between the asymptotic approximations. The additional variables observed in the high-dimensional asymptotics have zero factor loadings on the given factor, and thus do not provide any extra information.} 

There is also a related approach to factor models that allows for nonzero covariances between the errors. 
In the high-dimensional setting, this is handled by assuming an approximate factor structure, as in \cite{ChamberlainRothschild1983} or \cite{Bai2003}. 
In the low-dimensional setting, \cite{Williams2020} presents strategies for identifying factor model parameters using only a subset of the covariance restrictions. 
\cite{Williams2020} does not consider weak identification or identification-robust inference. 
Similar to \cite{Williams2020}, we can allow some nonzero covariances between the error terms. 

The remainder of the paper proceeds as follows. 
Section 2 describes a general low-dimensional factor model and presents options for identification-robust inference. 
Section 3 gives the reparameterization for a factor model with one factor. 
Section 4 gives the reparameterization for a factor model with two factors. 
Section 5 presents the simulations. 
Section 6 presents the empirical application. 
Section 7 concludes. 
Supplemental Materials contain additional details on the reparameterizations, simulations, and empirical application. 

\section{A Low-Dimensional Factor Model}
\label{Section2}

Suppose a researcher observes $p$ variables in a dataset, $W_i=(W_{1i}, W_{2i}, ... W_{pi})'$, for $i\in\{1,...,n\}$. 
A factor model for $W_i$ hypothesizes a common variable or factor that contributes to the variation of multiple $W_{ji}$. 
The factor model is defined by the equation, 
\begin{equation}
W_i=\Lambda f_i+\epsilon_i, \label{model}
\end{equation}
where $\Lambda$ is a $p\times m$ matrix of factor loadings, $f_i$ is an $m$-vector of unobserved factors, and $\epsilon_i$ is a $p$-vector of unobserved errors. 
In low-dimensional factor models, $p$ is fixed as the sample size increases. 
The model in (\ref{model}) can be generalized to allow $W_i$ to be the errors from a system of linear regressions; see Section A.2 in the Supplemental Materials. 
This includes linear regressions on a constant, so we can assume without loss of generality that $W_i$, $\epsilon_i$, and $f_i$ have mean zero. 

Let $\Sigma$ denote the covariance matrix of the factors, assumed to be positive definite, and let $\Phi$ denote the covariance matrix of the errors. 
We assume $\Phi$ is a diagonal matrix and that $f_i$ is uncorrelated with $\epsilon_i$. 
These assumptions ensure $f_i$ is the only source of common variation. 
Section A.3, in the Supplemental Materials, allows for some nonzero off-diagonal elements in $\Phi$. 

Factor models imply a covariance matrix relationship between the factors and the observed variables. 
If we let $\Omega$ denote the covariance matrix of $W_i$, then 
\begin{equation}
\Omega=\Lambda\Sigma\Lambda'+\Phi. \label{covariance_equation}
\end{equation}
This equation relates the factor-model parameters that we are interested in ($\Lambda$, $\Sigma$, and $\Phi$) to the covariance matrix of the observed variables. 
Identification and estimation of factor models is focused on exploiting this relationship. 

For identification, notice that (\ref{covariance_equation}) implies an indeterminacy in the $\Sigma$ and $\Lambda$ parameters. 
For any $m\times m$ invertible matrix, $M$, both $(\Lambda,\Sigma)$ and $(M\Lambda,M\inv\Sigma M\inv)$ imply the same value of $\Omega$. 
This means that $m^2$ additional restrictions are needed to identify $\Lambda$ and $\Sigma$ from (\ref{covariance_equation}). 
\cite{BaiLi2012} describe five sets of restrictions that are commonly used. 
In this paper, we assume the first $m$ rows of $\Lambda$ are $I_m$, the $m$-dimensional identity matrix. 
This corresponds to IC1 in \cite{BaiLi2012}. 
The weak identification analysis is the same under different restrictions, once the factors are appropriately rotated and rescaled; see Section A.4 in the Supplemental Materials. 
Other requirements for identification using (\ref{covariance_equation}) are discussed in Sections \ref{Section3} and \ref{Section4}. 

For estimation, (\ref{covariance_equation}) can be used as moments in a GMM model. 
Let $vec(A)$ denote the vectorization operator for a matrix $A$ and let $vech(A)$ denote the vectorization operator for a square symmetric matrix $A$ that only takes the values at or below the diagonal. 
Collect the factor model parameters as $\gamma=(vec(\Lambda)',vech(\Sigma)',diag(\Phi)')'$, and let $\Omega(\gamma)=\Lambda\Sigma\Lambda'+\Phi$.
We can write the moment function as $g(\gamma,W_i)=vech(W_iW'_i-\Omega(\gamma))$. 
The GMM estimator then minimizes 
\begin{equation}
Q_n(\gamma)=n\bar g_n(\gamma)'\widehat V^{-1}_n\bar g_n(\gamma), \label{GMM_objective}
\end{equation}
where $\bar g_n(\gamma)=n\inv \isum g(\gamma,W_i)$ and $\widehat V_n$ is a consistent estimator of the asymptotic variance of $\bar g_n(\gamma)$. 
Note that dependence in the data is allowed as long as $\bar g_n(\gamma)$ is asymptotically normal and $\widehat V_n$ is suitably chosen to be consistent for the asymptotic variance of $\bar g_n(\gamma)$. 
Using (\ref{covariance_equation}) as moments in a GMM model allows us to apply the weak identification theory that has been developed for GMM. 

Weak identification theory for GMM models provides several hypothesis tests that are identification-robust. 
In the rest of this section, we describe the main identification-robust hypothesis tests developed for GMM models. 
For the full-vector hypothesis, $H_0: \gamma=\gamma_0$, there are three main statistics.\footnote{Other statistics for the full-vector hypothesis include the empirical likelihood versions of the AR, K, and CLR statistics in \cite{GuggenbergerSmith2005} and \cite{GuggenbergerRamalhoSmith2012} and conditional linear combination statistics in \cite{Andrews2016}. In addition, \cite{AndrewsMikusheva2016Functional} propose a way to make other statistics identification-robust by conditioning on a sufficient statistic for a functional nuisance parameter.} 
\cite{StockWright2000} show that the Anderson-Rubin (AR) statistic, $Q_n(\gamma_0)$, can be compared to $\chi^2_{1-\alpha,k}$, where $\chi^2_{1-\alpha,k}$ denotes the $1-\alpha$ quantile of the chi-squared distribution with $k$ degrees of freedom and $k=p(p+1)/2$ is the number of moments. 
\cite{Kleibergen2005} shows that the K statistic, 
\begin{equation}
K(\gamma_0)=n\bar g_n(\gamma_0)' \widehat V_n^{-1/2} P_{\widehat V_n^{-1/2} \widehat D_n(\gamma_0)}\widehat V_n^{-1/2}\bar g_n(\gamma_0), 
\end{equation}
can be compared to $\chi^2_{1-\alpha, q}$, where $\widehat D_n(\gamma)=\frac{\partial}{\partial\gamma'}\bar g_n(\gamma)$, $q=(p-m)m+m(m+1)/2+p$ is the number of parameters, $P_C$ denotes the matrix that projects onto the span of the columns of $C$, and $\widehat V^{-1/2}_n$ denotes the inverse of the symmetric matrix square root of $\widehat V_n$. 
(This formula exploits the fact that the moments in (\ref{GMM_objective}) are additively separable, and thus the formula for $\widehat D_n(\gamma)$ simplifies.) 
\cite{Kleibergen2005} also shows that the conditional likelihood ratio (CLR) statistic, 
\begin{equation}
CLR(\gamma_0)=\left(Q_n(\gamma_0)-rk(\gamma_0)+\sqrt{(Q_n(\gamma_0)+rk(\gamma_0))^2-4J(\gamma_0)rk(\gamma_0)}\right)/2, 
\end{equation}
where $J(\gamma_0)=Q_n(\gamma_0)-K(\gamma_0)$ and $rk(\gamma_0)$ is a rank statistic satisfying certain conditions (see \cite{Kleibergen2005}), 
can be compared to the $1-\alpha$ quantile of the conditional distribution of $\left(X_1+X_2-rk(\gamma_0)+\sqrt{(X_1+X_2+rk(\gamma_0))^2-4X_2 rk(\gamma_0)}\right)/2$ given $rk(\gamma_0)$, where $X_1$ and $X_2$ are independent random variables with distributions $\chi^2_{q}$ and $\chi^2_{k-q}$, respectively. 

Next consider a subvector hypothesis, $H_0: \mu=\mu_0$, where $\gamma=(\mu,\nu)$ with $\mu$ denoting the parameter of interest (assumed to be a scalar for simplicity; a vector parameter of interest can be handled easily) and $\nu$ denoting the nuisance parameters. 
(1) The first approach is to project the full-vector tests; see \cite{Dufour1997} and \cite{DufourJasiak2001}. 
We reject $H_0: \mu=\mu_0$ if for every value of $\nu_0$, the hypothesis $H_0: \gamma=(\mu_0,\nu_0)$ is rejected using an identification-robust full-vector test. 
(2) The second approach is to plug-in an estimator for $\nu$. 
Let $\hat\nu_0$ denote the null-imposed estimator for $\nu$ that minimizes $Q_n(\mu_0,\nu)$. 
The plug-in version of the AR test rejects if $Q_n(\mu_0,\hat\nu_0)$ exceeds $\chi^2_{1-\alpha,k-q+1}$. 
The plug-in version of the K test rejects if $K(\mu_0,\hat\nu_0)$ exceeds $\chi^2_{1-\alpha,1}$. 
The plug-in version of the CLR test rejects if $\left(Q_n(\mu_0,\hat\nu_0)\hspace{-1mm}-\hspace{-1mm}rk(\mu_0)\hspace{-1mm}+\hspace{-1mm}\sqrt{(Q_n(\mu_0,\hat\nu_0)+rk(\mu_0))^2-4(Q_n(\mu_0,\hat\nu_0)\hspace{-1mm}-\hspace{-1mm}K(\mu_0,\hat\nu_0))rk(\mu_0)}\right)/2$, where $rk(\mu_0)$ is a rank statistic satisfying certain conditions, exceeds the $1-\alpha$ quantile of the conditional distribution of $\left(Y_1+Y_2-rk(\mu_0)+\sqrt{(Y_1+Y_2+rk(\mu_0))^2-4Y_2rk(\mu_0)}\right)/2$ given $rk(\mu_0)$, where $Y_1$ and $Y_2$ are independent random variables with distributions $\chi^2_1$ and $\chi^2_{k-q}$, respectively. 
The plug-in versions of the identification-robust subvector hypothesis tests are less conservative and more powerful because the degree of freedom in the critical value is reduced by the number of nuisance parameters that are plugged in. 
However, the plug-in versions of the identification-robust subvector hypothesis tests have only been shown to be valid under the assumption that the nuisance parameters are strongly identified. 
(3) The third approach is a two-step test for $H_0: \mu=\mu_0$. 
The first step calculates an identification-robust confidence set for $\nu$ with coverage probability $1-\alpha_1$. 
The second step projects a size-$\alpha_2$ full-vector test for $H_0: \gamma=(\mu_0,\nu)$ over all values of $\nu$ in the first-step confidence set, where $\alpha=\alpha_1+\alpha_2$. 
The two-step approach is similar to the projection approach, except that not all values of $\nu$ need to be projected over---only the values in the first-step confidence set. 
The two-step approach is useful because it does not require the nuisance parameters to be strongly identified. 
\cite{ChaudhuriZivot2011}, \cite{DAndrews2017}, and \cite{IAndrews2018} provide various implementations of the two-step approach. 

\textbf{Remarks.} \textbf{2.1.} The reparameterizations proposed in this paper allow us to apply existing theory for plug-in identification-robust subvector inference. 
The reparameterizations modify the nuisance parameters so they can be classified as strongly or weakly identified. 
In the original parameterization, the nuisance parameters cannot be so classified. 

\textbf{2.2.} \cite{Cox_weak_id_w_bounds} emphasizes an implicit requirement for the plug-in approach, which is that the parameter space for $\gamma$ must be a product space between the weakly identified parameters and the strongly identified parameters.\footnote{This comes from the fact that the plug-in approach requires $\hat\nu_0$ to be asymptotically normal. In particular, $\nu$ cannot be on the boundary of the null-imposed parameter space. At the same time, we want to allow $\mu_0$ to be on the boundary of its parameter space because that coincides with the extreme points of the identified set. If the parameter space is not a product space, then for some $\mu_0$ (on the boundary of the parameter space), $\nu$ will be on the boundary of the null-imposed parameter space.} 
In our implementation of the plug-in hypothesis tests, we are careful to satisfy this requirement. 

\textbf{2.3.} The assumption that $\Sigma$ is positive definite is also required in strongly identified factor models.\footnote{A factor model with fewer factors can be embedded within a factor model with more factors by setting some of the factor loadings to zero, so the assumption that $\Sigma$ is positive definite is innocuous.} 
If we imagine that the dataset includes the factors, so that (\ref{model}) can be viewed as a system of linear regressions, then the assumption that $\Sigma$ is positive definite ensures no multicollinearity in the factors. 
The validity of the reparameterizations proposed in this paper does not require the variances in $\Sigma$ to be bounded away from zero. 
However, placing such a bound is helpful for two reasons: (a) numerically minimizing (\ref{GMM_objective}) and (b) establishing uniform validity of the identification-robust hypothesis tests. 
We therefore include such a bound in the parameter spaces for the simulations and empirical application. 

\textbf{2.4.} We note that the maximum likelihood (ML) estimator is a more common estimator for low-dimensional factor models, where the factors and errors are assumed to be jointly normally distributed. 
The GMM estimator is asymptotically equivalent to ML under strong identification if the GMM objective function is efficiently weighted and the model is correctly specified. 
\cite{Cox_weak_id_w_bounds} considers weak identification in a class of minimum distance models that covers the ML estimator for low-dimensional factor models. 
Another estimator, based on a instrumental variables (IV) regression, is discussed in Section A.5 in the Supplemental Materials. 
\qed

\section{One Factor}
\label{Section3}

With one factor and $p$ observed variables, there are $2p$ parameters in the model. 
They are the variance of the factor, $\sigma^2$, the variances of the errors, $\Phi=\text{diag}(\phi_1,\phi_2,...,\phi_p)$, and $p-1$ factor loadings. 
The factor loading matrix is 
\begin{equation}
\Lambda = \left[\begin{array}{c}1\\\lambda_2\\\lambda_3\\\vdots\\\lambda_p\end{array}\right]. 
\end{equation}
Note that we use $\lambda_1=1$ as the additional restriction. 
It assigns the units of the factor to be the same as the units of $W_{1i}$. 

Identification is determined by equation (\ref{covariance_equation}), which writes the covariance matrix of $W_i$ as a nonlinear function of the parameters. 
Let $\omega_j=\text{Var}(W_{ji})$ for $j\in\{1,2,...,p\}$ and $\rho_j=\text{Cov}(W_{1i},W_{ji})$, for $j\in\{2,3,...,p\}$. 
In addition, let $\tau_{jk}=\text{Cov}(W_{ji},W_{ki})$ for $j\in\{2,...,p-1\}$ and $k\in\{j+1,...,p\}$. 
Equation (\ref{covariance_equation}) is composed of $p(p+1)/2$ scalar equations that set a nonlinear function of $2p$ factor model parameters equal to $p(p+1)/2$ identified variances/covariances of $W_i$. 
The factor model parameters are identified if and only if these equations can be inverted for a unique value of $\gamma$. 
\cite{AndersonRubin1956} give a simple necessary and sufficient condition on the factor loadings for $\gamma$ to be identified: at least two of $\lambda_2,...,\lambda_p$ need to be nonzero. 

We demonstrate the identification problem with $p=3$. 
We can write out equation (\ref{covariance_equation}) as 
\begin{equation}
\left[\begin{array}{ccc}\omega_1&\rho_2&\rho_3\\\rho_2&\omega_2&\tau_{23}\\\rho_3&\tau_{23}&\omega_3\end{array}\right]=\left[\begin{array}{ccc}\sigma^2+\phi_1&\lambda_2\sigma^2&\lambda_3\sigma^2\\\lambda_2\sigma^2&\lambda_2^2\sigma^2+\phi_2&\lambda_2\lambda_3\sigma^2\\\lambda_3\sigma^2&\lambda_2\lambda_3\sigma^2&\lambda_3^2\sigma^2+\phi_3\end{array}\right]. \label{Ex1_covariance_equation}
\end{equation}
Equation (\ref{Ex1_covariance_equation}) is composed of six scalar equations that set a nonlinear function of six unknown factor model parameters, $\gamma=(\lambda_2,\lambda_3,\sigma^2,\phi_1,\phi_2,\phi_3)'$, equal to six identified variances/covariances of $W_i$, $vech(\Omega)=(\omega_1,\rho_2,\rho_3,\omega_2,\tau_{23},\omega_3)'$. 
If both $\lambda_2\neq 0$ and $\lambda_3\neq 0$, then $\gamma$ is identified. 
When $\lambda_2\neq 0$ and $\lambda_3\neq 0$, then $\sigma^2$ can be identified by $\sigma^2=\tau_{23}\inv\rho_2\rho_3$. 
In this case, the other parameters in $\gamma$ are then easily identified. 
When $\lambda_3=0$ (or, by symmetry, $\lambda_2=0$), it is easy to see that $\gamma$ cannot be identified. 
The equations given by $\rho_3=\lambda_3\sigma^2$ and $\tau_{23}=\lambda_2\lambda_3\sigma^2$ are always zero. 
While these two equations can identify $\lambda_3=0$, there are still five other parameters to be identified and only four other equations. 
Thus, the identified set for $\gamma$ is given by a curve in the parameter space that satisfies $\lambda_3=0$ and these four other equations. 

When the identified set is determined by a curve in the parameter space, none of the parameters are strongly identified. 
This rules out the plug-in approach to identification-robust subvector inference. 
However, a careful reparameterization can change the nuisance parameters so that each parameter can be classified as weakly identified or strongly identified. 
\cite{HanMcCloskey2019} discuss this problem and provide a strategy for finding a reparameterization based on solving a sequence of differential equations using the Jacobian of the moments. 
Here, we give a closed-form reparameterization. 

\begin{reparameterization}\label{reparameterization1}
Let 
\begin{align*}
\rho_j&=\lambda_j\sigma^2 \text{ for }j\in\{2,...,p\}\\
\omega_j&=\lambda_j^2\sigma^2+\phi_j \text{ for }j\in\{1,...,p\}\\
\beta&=\sigma^2, 
\end{align*}
and let $\pi\hspace{-0.8mm}=\hspace{-0.8mm}(\rho',\omega')'$, where $\rho\hspace{-0.8mm}=\hspace{-0.8mm}(\rho_2,...,\rho_p)'$ and $\omega\hspace{-0.8mm}=\hspace{-0.8mm}(\omega_1,...,\omega_p)'$. 
The model in (\ref{model}) with one factor and $p$ observed variables can equivalently be parameterized by $\gamma=(\lambda_2,...,\lambda_p,\sigma^2,\phi_1,...,\phi_p)'$ or $\theta=(\pi',\beta)'$. 
\end{reparameterization}

\textbf{Remarks.} \textbf{3.1.} When $p=3$, Reparameterization \ref{reparameterization1} uses the scalar equations in (\ref{Ex1_covariance_equation}) to define the new parameters. 
The strategy is simple: replace an original parameter with a corresponding parameter from the covariance matrix of $W_i$. 
Note that we cannot replace $\sigma^2$ with $\tau_{23}=\lambda_2\lambda_3\sigma^2$ because that replacement would not be invertible when $\lambda_2\lambda_3=0$. 
Thus, we leave $\sigma^2$ in the parameters by simply redefining it to be $\beta$. 
The non-invertibility of $\tau_{23}$ for $\sigma^2$ becomes the key focus of the identification analysis. 

\textbf{3.2.} In the new parameterization, the strongly identified parameters are the ones in $\pi$, while the weakly identified parameter is $\beta$. 
Thus, each parameter can be classified as strongly or weakly identified. 

\textbf{3.3.} Estimation theory in weakly identified models requires further structure than just classifying the parameters; see Assumption C in \cite{StockWright2000}, Assumption A in \cite{AndrewsCheng2012}, and Assumption ID in \cite{HanMcCloskey2019}. 
The reparameterized factor model does not satisfy this further structure. 
Assumption C in \cite{StockWright2000} is very restrictive in nonlinear models; see the discussion in Section 2 in \cite{AndrewsGuggenberger2017}. 
Both Assumption A in \cite{AndrewsCheng2012} and Assumption ID in \cite{HanMcCloskey2019} require identification to be determined by whether a vector of strongly identified parameters is zero. 
Identification in the reparameterized factor model is determined by whether a \textit{function of} the strongly identified parameters is zero. 
It is unclear if there exists a further reparameterization that satisfies this further structure without adding assumptions to the model.\footnote{A natural attempt to further reparameterize the model involves defining a new parameter, $\nu=\rho_2\rho_3$, whose value determines identification. 
The problem is that $\nu$ is not invertible for $\rho_2$ when $\rho_3=0$ (and vice versa---$\nu$ is not invertible for $\rho_3$ when $\rho_2=0$). 
While this attempt does not work, it is hard to prove that such a further reparameterization does not exist. 
For this reason, we conclude that the existence of such a further reparameterization is ``unclear.''} 
The reparameterized model does satisfy the structure required by Theorem 1 in \cite{Cox_weak_id_w_bounds} when the objective function in (\ref{GMM_objective}) is recast as a minimum distance objective function. 
\qed \medskip

Using Reparameterization \ref{reparameterization1}, we can write the moments as functions of $\theta=(\pi',\beta)'$. 
There are three kinds of moments. 
For $j\in\{1,...,p\}$, the moments associated with the variance of $W_{ji}$ are 
\begin{equation}
g_{1j}(\theta,W_i)=W_{ji}^2-\omega_j. \label{first_type}
\end{equation}
For $j\in\{2,...,p\}$, the moments associated with the covariance between $W_{1i}$ and $W_{ji}$ are 
\begin{equation}
g_{2j}(\theta,W_i)=W_{1i}W_{ji}-\rho_j. 
\end{equation}
For $j\in\{2,...,p-1\}$ and $k\in\{j+1,...,p\}$, the moments associated with the covariance between $W_{ji}$ and $W_{ki}$ are 
\begin{equation}
g_{3jk}(\theta,W_i)=W_{ji}W_{ki}-\beta\inv\rho_j\rho_k. \label{third_type}
\end{equation}
The nontrivial moments are the third type. 
If any one of them can be inverted for $\beta$, then $\beta$ is identified. 
Otherwise, $\beta$ is not identified. 
Another way to say this is to let $s(\rho)$ denote the $(p-1)(p-2)/2$ dimensional vector composed of $\rho_j\rho_k$ for $j\in\{2,...,p-1\}$ and $k\in\{j+1,...,p\}$ and evaluate whether $s(\rho)$ is equal to zero. 
Note that this condition is equivalent to the \cite{AndersonRubin1956} condition requiring $\lambda_j$ and $\lambda_k$ to be nonzero for some $j\neq k\in\{2,...,p\}$.\footnote{This demonstrates an important part of the classification needed for weak identification in GMM models. The identification status of potentially weakly identified parameters must be determined only by strongly identified parameters. According to this principle, the condition from \cite{AndersonRubin1956} is insufficient because the factor loadings are not strongly identified. Reparameterization \ref{reparameterization1} is careful to translate this condition to an equivalent one on the new parameters.} 

Weak identification arises when the true value of the parameters is considered as a sequence of values indexed by the sample size that converges to a non-identified limit at a particular rate. 
Denote this sequence of true values by a subscript $n$. 
To determine the correct rate, we look at the columns of the Jacobian of the moments. 
When $p=3$, we can write out the Jacobian: 
\begin{equation}
\frac{\partial}{\partial\theta'}g(\theta,W_i)=G(\theta)=-\left[\begin{array}{cccccc}0&0&1&0&0&0\\1&0&0&0&0&0\\0&1&0&0&0&0\\0&0&0&1&0&0\\\beta\inv\rho_3&\beta\inv\rho_2&0&0&0&-\beta^{-2}\rho_2\rho_3\\0&0&0&0&1&0\end{array}\right]. 
\end{equation}
The only column that can go to zero is the last one, which depends on $-\beta^{-2}\rho_2\rho_3$. 
If this value converges to zero at the $n^{-1/2}$ rate, then we have weak identification. 

When $p>3$, the column of the Jacobian of the third type of moments in (\ref{third_type}) associated with the derivative with respect to $\beta$ is proportional to $s(\rho)$. 
When we consider a sequence of true values of $\rho$, indexed by $n$, weak identification arises when $s(\rho_n)$ converges to zero at the $n^{-1/2}$ rate. 

\begin{definition}\label{def1}
A sequence of parameters, $\theta_n=(\pi'_n,\beta_n)'=(\rho_n,\omega_n,\beta_n)'$, induces weak identification of the model in (\ref{model}) with one factor and $p$ observed variables if $\sqrt{n}s(\rho_n)\rightarrow b$ for some $b\in\R^{(p-1)(p-2)/2}$. 
\end{definition}

\textbf{Remarks.} \textbf{3.4.} In Definition \ref{def1}, the key values that determine the strength of identification are the products between $\rho_{jn}$ and $\rho_{kn}$. 
These products can converge to zero at the $n^{-1/2}$ rate in a variety of ways. 
One simple way is if all but one value of $\rho_{jn}$ converges to zero at the $n^{-1/2}$ rate. 
Another simple way is if all the $\rho_{jn}$ values converge to zero at the $n^{-1/4}$ rate. 
The practical consequence of this slower convergence rate is that the influence of weak identification covers a larger neighborhood of $\rho=0$ in the parameter space---one that shrinks slower than $n^{-1/2}$. 

\textbf{3.5.} While a formal test for weak identification is beyond the scope of the paper, Definition \ref{def1} suggests detecting weak identification should be based on $s(\rho)$ because $\rho$ is strongly identified. 
(The elements of $\rho$ are covariances between $W_{ji}$ and $W_{ki}$ for $j, k\in\{2,...,p\}$ with $j\neq k$.) 
Conditions for identification that depend on potentially weakly identified parameters, such as factor loadings or signal-to-noise ratios, are less reliable for detecting weak identification. 
\qed\medskip

Reparameterization \ref{reparameterization1} is useful for identification-robust hypothesis testing.\footnote{By inverting identification-robust hypothesis tests, Reparameterization \ref{reparameterization1} is also useful for constructing identification-robust confidence intervals.} 
Suppose we are interested in testing $H_0: r(\pi,\beta)=r_0$, for some function $r(\pi,\beta)$ and some hypothesized value $r_0$. 
There are two approaches that we can follow, depending on whether $r(\pi,\beta)$ can be inverted for $\beta$. 

(A) The first approach applies when $r(\pi,\beta)$ can be inverted for $\beta$.\footnote{We have to assume that $r(\pi,\beta)$ is invertible for one of its arguments because weak identification theory for the plug-in approach only covers subvector inference, and not inference on a general function of the parameters; see Section 3.3 in \cite{Kleibergen2005}. When $r(\pi,\beta)$ takes vector values, then we assume it is invertible for a subvector of its arguments that includes $\beta$.} 
We further reparameterize the model using $r=r(\pi,\beta)$. 
That is, the model can be equivalently parameterized by $\theta=(\pi,\beta)$ or $\ddot\theta=(\pi,r)$. 
In this case, the parameter being tested is the potentially weakly identified parameter, and all the nuisance parameters are strongly identified. 
We use the inverse of $r(\pi,\beta)$ for $\beta$ to rewrite the moments in (\ref{first_type})-(\ref{third_type}) as functions of $\ddot\theta$. 
Then, we test $H_0$ using a version of the AR, K, or CLR tests that plugs in the nuisance parameters. 

(B) The second approach applies when $r(\pi,\beta)$ cannot be inverted for $\beta$, but it can be inverted for one of the parameters in $\pi$. 
Let the new parameterization be given by $\ddot\theta=(\tilde\pi,r,\beta)$, where $\tilde\pi$ is a subvector of $\pi$. 
In this case, we can test $H_0$ using a version of the AR, K, or CLR tests that plugs in the strongly identified nuisance parameters and projects only over the potentially weakly identified one. 
That is, for any $\beta_0$, we consider the joint hypothesis $\widetilde H_0: r=r_0$ and $\beta=\beta_0$. 
All the nuisance parameters are strongly identified, so we can test $\widetilde H_0$ using a plug-in test. 
We reject $H_0$ if $\widetilde H_0$ rejects for all values of $\beta_0$. 

\begin{table}[p]
\begin{doublespacing}
\begin{center}
\scalebox{\shrinkageparameter}{
\begin{threeparttable}
{\scriptsize
\caption{Invertible Hypotheses in a One-Factor Model}\label{table-1F-Hypotheses}
\begin{center}
\begin{tabular}{llc}
\hline\hline\vspace{-0.2cm}&&\\
Hypothesis&Further Reparameterization&Additional Assumption\\
\hline\vspace{-0.2cm}\\
FV&$r(\pi,\beta)=\beta$&-\\
$j$th FL&$r(\pi,\beta)=\beta\inv\rho_j$&$\lambda_j\neq 0$\\
First EV&$r(\pi,\beta)=\omega_1-\beta$&-\\
$j$th EV&$r(\pi,\beta)=\omega_j-\beta^{-1}\rho_j^2$&$\lambda_j\neq 0$\\
First StNR&$r(\pi,\beta)=(\omega_1-\beta)\inv\beta$&-\\
$j$th StNR&$r(\pi,\beta)=(\beta\omega_j-\rho_j^2)\inv\rho_j^2$&$\lambda_j\neq 0$\\
\hline
\end{tabular}
{\singlespace
\begin{tablenotes}
\item {\em Note:} The hypothesis is $H_0: r(\pi,\beta)=r_0$, where $r(\pi,\beta)$ is given in the second column, and $r_0$ is a hypothesized value. ``First'' indicates that the hypothesis corresponds to the first observed variable, $W_{1i}$, while ``$j$th'' indicates that the hypothesis corresponds to the $j$th observed variable, $W_{ji}$, for $j\in\{2,...,p\}$. ``FV'' stands for ``Factor Variance'', ``FL'' stands for ``Factor Loading'', ``EV'' stands for ``Error Variance'', and ``StNR'' stands for ``Signal to Noise Ratio''. The ``-'' indicates that no additional assumption is needed for $r(\pi,\beta)$ to be invertible for $\beta$. 
\end{tablenotes}
}
\end{center}
}
\end{threeparttable}
}
\end{center}
\end{doublespacing}
\end{table}

Suppose the researcher is interested in testing the value of the variance of the factor. 
In this case, $r(\pi,\beta)=\beta$ is trivially invertible for $\beta$. 
We can follow the first approach and plug in all the nuisance parameters. 
As another example, suppose one is interested in testing the value of the $j$th factor loading. 
In this case, $r(\pi,\beta)=\beta^{-1}\rho_j$. 
This is not invertible for $\beta$ because $\rho_j$ may be zero. 
We can then follow the second approach and invert $r(\pi,\beta)$ for $\rho_j$. 
While $\beta$ is a nuisance parameter that is weakly identified, all the other nuisance parameters can be plugged in because they are strongly identified. 
The disadvantage of the second approach is that we have to project over the nuisance parameter, which, in practice, means that we need to add a degree of freedom to the chi-squared distribution used to calculate the critical value for the plug-in AR, K, or CLR tests. 

In some cases, relatively simple assumptions can be added to ensure $r(\pi,\beta)$ is invertible for $\beta$. 
For example, if we add the assumption that $\lambda_j\neq 0$ (and therefore $\rho_j\neq 0$), then $r(\pi,\beta)=\beta^{-1}\rho_j$ is invertible for $\beta$. 
In that case, we can follow the first approach and plug in all the nuisance parameters. 
Table \ref{table-1F-Hypotheses} lists some hypotheses in the one-factor model that can be inverted for $\beta$, so the first approach applies, and any additional assumptions required. 
Current empirical practice includes reporting estimates of factor loadings, the variance of the factor, and signal to noise ratios, together with confidence intervals. 
The two plug-in approaches described in this section provide identification-robust versions of the confidence intervals. 
Table \ref{table-1F-Hypotheses} covers these commonly reported parameters. 

\textbf{Remarks.} \textbf{3.6.}
Note the value of Reparameterization \ref{reparameterization1}: when using the first approach, we do not need to project at all, and when using the second approach, we only need to project over one nuisance parameter. 
With one factor and $p$ observed variables, there are $2p-1$ nuisance parameters. 
The value of the reparameterization increases as the number of observed variables increases because the number of nuisance parameters also increases. 

\textbf{3.7.} The plug-in AR test is particularly convenient for the second approach because projecting over $\beta$ does not require a grid of values for $\beta$. 
Because the AR statistic is just the value of the GMM objective function, the AR test rejects $\widetilde H_0$ for all values of $\beta_0$ if it rejects for the value of $\beta_0$ that minimizes the GMM objective function under $H_0$. 
\qed

\section{Two Factors}
\label{Section4}

With two factors and $p$ observed variables, there are $3p-1$ parameters in the model. 
The covariance matrix of the factors, $\Sigma$, contains the variances of the factors, $\sigma^2_1$ and $\sigma^2_2$, and the covariance between the factors, $\sigma_{12}$. 
The variance matrix of the errors contains $p$ parameters, $\Phi=\text{diag}(\phi_1,\phi_2,...,\phi_p)$. 
The factor loading matrix contains $2p-4$ parameters: 
\begin{equation}
\Lambda = \left[\begin{array}{cc}1&0\\0&1\\\lambda_{31}&\lambda_{32}\\\vdots&\vdots\\\lambda_{p1}&\lambda_{p2}\end{array}\right]. 
\end{equation}
Note that we use $(\lambda_{11},\lambda_{12},\lambda_{21},\lambda_{22})=(1,0,0,1)$ as the additional restrictions. 
These restrictions assign the units of the factors to be the same as the units of $W_{1i}$ and $W_{2i}$. 
They also require $W_{1i}$ to not depend on the second factor and $W_{2i}$ to not depend on the first factor. 

In this paper, we assume $\lambda_{31}\neq 0$ and $\lambda_{41}\neq 0$. 
These assumptions can be interpreted by imagining a dataset that includes the factors and running a regression of the third/fourth variable on the factors. 
In that regression, the assumption that $\lambda_{31}\neq 0$ and $\lambda_{41}\neq 0$ are assumptions that the coefficients in the regression are nonzero. 

Identification is determined by equation (\ref{covariance_equation}), which writes the covariance matrix of $W_i$ as a nonlinear function of the parameters. 
Equation (\ref{covariance_equation}) is composed of $p(p+1)/2$ scalar equations that set a nonlinear function of $3p-1$ factor model parameters equal to $p(p+1)/2$ identified variances/covariances of $W_i$. 
The factor model parameters are identified if and only if these equations can be inverted for a unique value of $\gamma$. 
\cite{AndersonRubin1956} give a necessary and sufficient condition on the factor loadings for $\gamma$ to be identified from (\ref{covariance_equation}). 
If, for any row deleted from $\Lambda$, the remaining rows can be rearranged into two rank-2 matrices, then the factor model parameters are identified. 
Otherwise, they are not. 

The row-deletion condition is difficult to interpret. 
It is easier to think about some simple requirements for the row-deletion condition. 
The row-deletion condition requires (a) both factors have at least three nonzero factor loadings, (b) the factor loadings have at least five nonzero rows, and (c) the factor loadings for the non-normalized variables have rank 2: $(\lambda_{31},...,\lambda_{p1})$ is not collinear with $(\lambda_{32},...,\lambda_{p2})$. 
These are necessary conditions for identification. 
Thus, our assumptions are weaker than the assumptions required in an identified factor model. 

We next give the reparameterization that satisfies the weak identification classification in GMM models. 
We follow the same strategy as Reparameterization \ref{reparameterization1} and define the new parameters to be elements of the covariance matrix of $W_i$. 
Let $\lambda_j$ denote the $2\times 1$ vector denoting the $j$th row of $\Lambda$ for $j\in\{1,...,p\}$. 
Specifically, $\lambda_1=(1,0)'$ and $\lambda_2=(0,1)'$, because of the additional restrictions, while $\lambda_j=(\lambda_{j1},\lambda_{j2})'$ for $j\in\{3,...,p\}$. 

\begin{reparameterization}\label{reparameterization2}
Let 
\begin{alignat*}{3}
\omega_j&=Var(W_{ji})&&=\lambda'_j\Sigma\lambda_j+\phi_j \text{ for }j\in\{1,...,p\}\\
\rho_{j1}&=Cov(W_{ji},W_{1i})&&=\lambda'_1\Sigma\lambda_j \text{ for }j\in\{3,...,p\}\\
\rho_{j2}&=Cov(W_{ji},W_{2i})&&=\lambda'_2\Sigma\lambda_j \text{ for }j\in\{3,...,p\}\\
\chi&=Cov(W_{3i}, W_{4i})&&=\lambda'_3\Sigma\lambda_4. 
\end{alignat*}
Also let $\beta=\sigma^2_2$, $\omega=(\omega_1,...,\omega_p)'$, $\rho=(\rho_{31},\rho_{41},...,\rho_{p1},\rho_{32},\rho_{42},...,\rho_{p2})'$, and $\pi=(\rho',\omega',\chi,\sigma_{12})'$. 
The model in (\ref{model}) with two factors and $p$ observed variables can equivalently be parameterized by $\gamma=(vec(\Lambda)',vech(\Sigma)',diag(\Phi)')'$ or $\theta=(\pi',\beta)'$. 
\end{reparameterization}

\textbf{Remarks.} \textbf{4.1.} Reparameterization \ref{reparameterization2} uses the equations in (\ref{covariance_equation}) to define the new parameters. 
The strategy is the same as before: replace an original parameter with a corresponding identified parameter from the covariance matrix of $W_i$. 
Section A.6, in the Supplemental Materials, shows that this reparameterization is well-defined and invertible. 
$3p-2$ of the $3p-1$ parameters can be replaced in a way that is invertible. 
The remaining parameter is $\beta=\sigma^2_2$, the variance of the second factor. 
$\beta$ is potentially weakly identified depending on the value of $\pi$. 

\textbf{4.2.} As before, in the new parameterization, the identified parameters are the ones in $\pi$, while the weakly identified parameter is $\beta$. 
Thus, each parameter can be classified as strongly or weakly identified. 
\qed \medskip

Using Reparameterization \ref{reparameterization2}, we can write the moments as functions of $\theta=(\pi',\beta)'$. 
There are six kinds of moments. 
For $j\in\{1,...,p\}$, the moments associated with the variance of $W_{ji}$ are 
\begin{equation}
g_{1j}(\theta,W_i)=W_{ji}^2-\omega_j. 
\end{equation}
For $j\in\{3,...,p\}$, the moments associated with the covariance between $W_{1i}$ and $W_{ji}$ are 
\begin{equation}
g_{2j}(\theta,W_i)=W_{1i}W_{ji}-\rho_{j1}. 
\end{equation}
For $j\in\{3,...,p\}$, the moments associated with the covariance between $W_{2i}$ and $W_{ji}$ are 
\begin{equation}
g_{3j}(\theta,W_i)=W_{2i}W_{ji}-\rho_{j2}. 
\end{equation}
The moment associated with the covariance between $W_{1i}$ and $W_{2i}$ is 
\begin{equation}
g_{4}(\theta,W_i)=W_{1i}W_{2i}-\sigma_{12}. 
\end{equation}
The moment associated with the covariance between $W_{3i}$ and $W_{4i}$ is 
\begin{equation}
g_{5}(\theta,W_i)=W_{3i}W_{4i}-\chi. 
\end{equation}
For $j\in\{3,...,p-1\}$ and for $k\in\{max(j+1,5),...,p\}$, the moments associated with the covariance between $W_{ji}$ and $W_{ki}$ are 
\begin{equation}
g_{6jk}(\theta,W_i)=W_{ji}W_{ki}-\tau_{jk}(\theta), \label{equation_18}
\end{equation}
where $\tau_{jk}(\theta)$ is defined in the next paragraph. 

The nontrivial moments are the sixth type. 
For $j=3$ and $k\in\{5,...,p\}$, let 
\begin{equation}
\tau_{3k}(\pi,\beta)=\frac{\rho_{32}(\rho_{k2}\rho_{41}-\rho_{k1}\rho_{42})+\chi(\beta\rho_{k1}-\sigma_{12}\rho_{k2})}{\beta\rho_{41}-\sigma_{12}\rho_{42}}. 
\end{equation}
For $j=4$ and $k\in\{5,...,p\}$, let 
\begin{equation}
\tau_{4k}(\pi,\beta)=\frac{\rho_{42}(\rho_{k2}\rho_{31}-\rho_{k1}\rho_{32})+\chi(\beta\rho_{k1}-\sigma_{12}\rho_{k2})}{\beta\rho_{31}-\sigma_{12}\rho_{32}}. 
\end{equation}
For $j\in\{5,...,p-1\}$ and $k\in\{j+1,...,p\}$, let 
\begin{equation}
\tau_{jk}(\pi,\beta)=\frac{\beta q(\rho)+\sigma_{12}r(\rho)+\chi(\beta\rho_{k1}-\sigma_{12}\rho_{k2})(\beta\rho_{j1}-\sigma_{12}\rho_{j2})}{(\beta\rho_{31}-\sigma_{12}\rho_{32})(\beta\rho_{41}-\sigma_{12}\rho_{42})}, 
\end{equation}
where 
\begin{align*}
q(\rho)&=\rho_{31}\rho_{41}\rho_{j2}\rho_{k2}-\rho_{32}\rho_{42}\rho_{j1}\rho_{k1}\\
r(\rho)&=\rho_{32}\rho_{42}\rho_{j1}\rho_{k2}+\rho_{32}\rho_{42}\rho_{j2}\rho_{k1}-\rho_{31}\rho_{42}\rho_{j2}\rho_{k2}-\rho_{32}\rho_{41}\rho_{j2}\rho_{k2}.
\end{align*}
Note that the denominators in the above expressions are nonzero because $\beta\rho_{41}-\sigma_{12}\rho_{42}=(\sigma_1^2\sigma_2^2-\sigma_{12}^2)\lambda_{41}\neq 0$ and $\beta\rho_{31}-\sigma_{12}\rho_{32}=(\sigma_1^2\sigma_2^2-\sigma_{12}^2)\lambda_{31}\neq 0$.
If any one of these functions can be inverted for $\beta$, then $\beta$ is identified. 
The following proposition gives conditions for these functions to be invertible for $\beta$. 

\begin{proposition}\label{prop2}
Let $\pi=(\rho'_1,\rho'_2,\omega',\chi,\sigma_{12})'$, where $\rho_1=(\rho_{31},...,\rho_{p1})'$, $\rho_2=(\rho_{32},...,\rho_{p2})'$, and $\omega=(\omega_1,...,\omega_p)'$. 
For any $k\in\{5,...,p\}$, $\tau_{3k}(\pi,\beta)$ can be inverted for the value of $\beta$ if and only if $(\rho_{k2}\rho_{41}-\rho_{k1}\rho_{42})(\rho_{32}\rho_{41}-\chi\sigma_{12})\neq 0$. 
For any $k\in\{5,...,p\}$, $\tau_{4k}(\pi,\beta)$ can be inverted for the value of $\beta$ if and only if $(\rho_{k2}\rho_{31}-\rho_{k1}\rho_{32})(\rho_{31}\rho_{42}-\chi\sigma_{12})\neq 0$. 
Furthermore, for any $j\in\{5,...,p-1\}$ and $k\in\{j+1,...,p\}$, if we assume that $(\rho_{j2}\rho_{41}-\rho_{j1}\rho_{42})(\rho_{32}\rho_{41}-\chi\sigma_{12})=0$ and $(\rho_{k2}\rho_{31}-\rho_{k1}\rho_{32})(\rho_{31}\rho_{42}-\chi\sigma_{12})=0$, then $\tau_{jk}(\pi,\beta)$ can be inverted for the value of $\beta$ if and only if $(\rho_{j2}\rho_{41}-\rho_{j1}\rho_{42})(\rho_{k2}\rho_{31}-\rho_{k1}\rho_{32})\neq 0$. 
\end{proposition}

\textbf{Remarks.} \textbf{4.3.} When $p=5$, Proposition \ref{prop2} simplifies. 
In this case, there are no $\tau_{jk}(\pi,\beta)$ with $j\ge 5$. 
Let $s_1(\pi)=(\rho_{52}\rho_{41}-\rho_{51}\rho_{42})(\rho_{32}\rho_{41}-\chi\sigma_{12})$ and $s_2(\pi)=(\rho_{52}\rho_{31}-\rho_{51}\rho_{32})(\rho_{31}\rho_{42}-\chi\sigma_{12})$. 
Proposition \ref{prop2} says that $\beta$ is identified if and only if $(s_1(\pi),s_2(\pi))\neq (0,0)$. 

\textbf{4.4.} More generally, for $p>5$, let $s_1(\pi)$ be the $p-4$ dimensional vector composed of $(\rho_{k2}\rho_{41}-\rho_{k1}\rho_{42})(\rho_{32}\rho_{41}-\chi\sigma_{12})$ for $k\in\{5,...,p\}$. 
Let $s_2(\pi)$ be the $p-4$ dimensional vector composed of $(\rho_{k2}\rho_{31}-\rho_{k1}\rho_{32})(\rho_{31}\rho_{42}-\chi\sigma_{12})$ for $k\in\{5,...,p\}$. 
Also let $s_3(\pi)$ be the $(p-4)(p-5)/2$ dimensional vector composed of $(\rho_{j2}\rho_{41}-\rho_{j1}\rho_{42})(\rho_{k2}\rho_{31}-\rho_{k1}\rho_{32})$ for $j\in\{5,...,p-1\}$ and $k\in\{j+1,...,p\}$. 
Proposition \ref{prop2} says that $\beta$ is identified if and only if $(s_1(\pi),s_2(\pi),s_3(\pi))\neq (0,0,0)$. 

\textbf{4.5.} Notice that the $s(\pi)$ components are composed of relatively few subcomponents: $(\rho_{j2}\rho_{41}-\rho_{j1}\rho_{42})$, $(\rho_{k2}\rho_{31}-\rho_{k1}\rho_{32})$, $(\rho_{32}\rho_{41}-\chi\sigma_{12})$, and $(\rho_{31}\rho_{42}-\chi\sigma_{12})$. 
One could write down conditions for invertibility on the subcomponents, but they tend to break down into many cases. 
The conditions on the products in Proposition \ref{prop2} are simpler and are useful for characterizing weak identification, below. 
\qed\medskip

To define weak identification, we evaluate the derivative of $\tau_{jk}(\pi,\beta)$ with respect to $\beta$. 
When $p=5$, there are only two nontrivial moments, given by $\tau_{35}(\pi,\beta)$ and $\tau_{45}(\pi,\beta)$. 
We evaluate the derivative of these two moments with respect to $\beta$: 
\begin{equation}\label{jacobian2}
\frac{\partial}{\partial\beta}\left[\begin{array}{c}\tau_{35}(\pi,\beta)\\\tau_{45}(\pi,\beta)\end{array}\right]=\left[\begin{array}{c}s_1(\pi)(\rho_{41}\beta-\sigma_{12}\rho_{42})^{-2}\\s_2(\pi)(\rho_{31}\beta-\sigma_{12}\rho_{32})^{-2}\end{array}\right]. 
\end{equation}
Since $\rho_{41}\beta-\sigma_{12}\rho_{42}=(\sigma_1^2\sigma_2^2-\sigma_{12}^2)\lambda_{41}\neq 0$ and $\rho_{31}\beta-\sigma_{12}\rho_{32}=(\sigma_1^2\sigma_2^2-\sigma_{12}^2)\lambda_{31}\neq 0$, the only way (\ref{jacobian2}) can go to zero is if $s_1(\pi)$ and $s_2(\pi)$ converge to zero. 
Weak identification arises when these converge at the $n^{-1/2}$ rate. 

When $p>5$, let $s_{1,k}(\pi)$ denote the element of $s_1(\pi)$ corresponding to $\tau_{3k}(\pi,\beta)$, and similarly for $s_{2,k}(\pi)$ and $s_{3, jk}(\pi)$. 
The derivatives for $\tau_{3k}(\pi,\beta)$ and $\tau_{4k}(\pi,\beta)$ are 
\begin{equation}\label{jacobian3}
\frac{\partial}{\partial\beta}
\left[\begin{array}{c}
\tau_{3k}(\pi,\beta)\\\tau_{4k}(\pi,\beta)
\end{array}\right]
=\left[\begin{array}{c}
s_{1,k}(\pi)(\rho_{41}\beta-\sigma_{12}\rho_{42})^{-2}\\s_{2,k}(\pi)(\rho_{31}\beta-\sigma_{12}\rho_{32})^{-2}
\end{array}\right],  
\end{equation}
for any $k\in\{5,...,p\}$. 
The derivatives for $\tau_{jk}(\pi,\beta)$ are somewhat more complicated: 
\begin{align}
\frac{\partial}{\partial\beta} \tau_{jk}(\pi,\beta)
=&\frac{-s_{1,j}(\pi)(\beta\rho_{k1}-\sigma_{12}\rho_{k2})}{(\rho_{41}\beta-\sigma_{12}\rho_{42})^{2}(\rho_{31}\beta-\sigma_{12}\rho_{32})}\label{jacobian4}\\
&-\frac{s_{2,k}(\pi)(\beta\rho_{j1}-\sigma_{12}\rho_{j2})}{(\rho_{41}\beta-\sigma_{12}\rho_{42})(\rho_{31}\beta-\sigma_{12}\rho_{32})^{2}}\nonumber\\
&-\frac{s_{3, jk}(\pi)}{(\rho_{41}\beta-\sigma_{12}\rho_{42})^{2}(\rho_{31}\beta-\sigma_{12}\rho_{32})^{2}}\nonumber
\end{align}
for any $j\in\{5,...,p-1\}$ and $k\in\{j+1,...,p\}$. 
Since $\rho_{41}\beta-\sigma_{12}\rho_{42}=(\sigma_1^2\sigma_2^2-\sigma_{12}^2)\lambda_{41}\neq 0$ and $\rho_{31}\beta-\sigma_{12}\rho_{32}=(\sigma_1^2\sigma_2^2-\sigma_{12}^2)\lambda_{31}\neq 0$, the only way (\ref{jacobian3}) can go to zero is if $s_{1,k}(\pi)$ and $s_{2,k}(\pi)$ both converge to zero. 
Weak identification requires these to converge at the $n^{-1/2}$ rate. 
Once $s_{1,j}(\pi)$ and $s_{2,k}(\pi)$ converge at the $n^{-1/2}$ rate, then the first two terms in (\ref{jacobian4}) converge at least at the $n^{-1/2}$ rate.\footnote{The other numerator terms, $\beta\rho_{k1}-\sigma_{12}\rho_{k2}=(\sigma_1^2\sigma_2^2-\sigma_{12}^2)\lambda_{k1}$ and $\beta\rho_{j1}-\sigma_{12}\rho_{j2}=(\sigma_1^2\sigma_2^2-\sigma_{12}^2)\lambda_{j1}$, are assumed to converge to limiting parameter values, and thus are bounded. They may converge to zero, and then the rate would be faster.} 
Therefore, the expression in (\ref{jacobian4}) converges at the $n^{-1/2}$ rate if and only if $s_{3,jk}(\pi)$ converges at the $n^{-1/2}$ rate. 
We use this to define weak identification. 

\begin{definition}\label{def2}
A sequence of parameters, $\theta_n=(\pi'_n,\beta_n)'$, induces weak identification of the model in (\ref{model}) with two factors and five observed variables if $\sqrt{n}s_1(\pi_n)\rightarrow b_1$ and $\sqrt{n}s_2(\pi_n)\rightarrow b_2$ for some $(b_1,b_2)\in\R^2$. 

A sequence of parameters, $\theta_n=(\pi'_n,\beta_n)'$, induces weak identification of the model in (\ref{model}) with two factors and $p>5$ observed variables if $\sqrt{n}(s_1(\pi_n),s_2(\pi_n),s_3(\pi_n))$ $\rightarrow b$ for some $b\in\R^{(p-2)(p-3)/2-1}$. 
\end{definition}

\textbf{Remarks.} \textbf{4.6.} In Definition \ref{def2}, the key values that determine the strength of identification are $s_1(\pi)$, $s_2(\pi)$, and $s_3(\pi)$. 
All three must be converging to zero at the $n^{-1/2}$ rate for weak identification. 
Notice that $s_1(\pi)$, $s_2(\pi)$, and $s_3(\pi)$ are themselves products of components, each of which can go to zero. 
There are lots of possible sequences of parameters $\pi_n$ converging at various rates to a limit that induce $n^{-1/2}$ convergence of $s(\pi)=(s_1(\pi),s_2(\pi),s_3(\pi))$. 
For example, if $p=5$, then $\rho_{52}\rho_{41}-\rho_{51}\rho_{42}$ can converge to zero at the $n^{-1/4}$ rate while $\rho_{41}\rho_{32}-\chi\sigma_{12}$ can converge to zero at the $n^{-1/4}$ rate to get convergence of $s_1(\pi_n)$ at the $n^{-1/2}$ rate. 

\textbf{4.7.} This characterization of weak identification is more useful for us than the row-deletion condition from \cite{AndersonRubin1956} because it is stated in terms of strongly identified parameters. 
It suggests detecting weak identification should be based on $s(\pi)$. 
\qed\medskip

Reparameterization \ref{reparameterization2} is useful for identification-robust hypothesis testing. 
Suppose we are interested in testing $H_0: r(\pi,\beta)=r_0$, for some function $r(\pi,\beta)$ and some hypothesized value, $r_0$. 
Just as in the one-factor model, there are two approaches we can follow, depending on whether $r(\pi,\beta)$ can be inverted for $\beta$. 

Suppose the researcher is interested in testing the variance of the second factor. 
In this case, $r(\pi,\beta)=\beta$ is trivially invertible for $\beta$. 
We can follow the first approach and plug in all the nuisance parameters. 
If, however, the researcher is interested in one of the factor loadings, say $\lambda_{31}$, then $r(\pi,\beta)=\lambda_{31}(\pi,\beta)=\frac{\chi\beta-\rho_{42}\rho_{32}}{\beta\rho_{41}-\sigma_{12}\rho_{42}}$, where $\lambda_{31}(\pi,\beta)$ is the inverse of Reparameterization \ref{reparameterization2} for $\lambda_{31}$. 
This hypothesis is not invertible for $\beta$ in general. 
We then follow the second approach and plug in all the nuisance parameters except for $\beta$. (In general, $\lambda_{31}(\pi,\beta)$ is invertible for $\chi$.) 
If we add the assumption that $\lambda_{32}\neq 0$ and $\rho_{42}\neq 0$, then some algebra shows that $r(\pi,\beta)$ can be inverted for $\beta$.\footnote{If $\hat\beta$ and $\tilde\beta$ are two different values of $\beta$, then some algebra shows that $\lambda_{31}(\pi,\hat\beta)-\lambda_{31}(\pi,\tilde\beta)=\frac{(\hat\beta-\tilde\beta)\rho_{42}(\rho_{41}\rho_{32}-\chi\sigma_{12})}{(\hat\beta\rho_{41}-\sigma_{12}\rho_{42})(\tilde\beta\rho_{41}-\sigma_{12}\rho_{42})}\neq 0$, where $\rho_{41}\rho_{32}-\chi\sigma_{12}=\lambda_{41}\lambda_{32}(\sigma_1^2\sigma_2^2-\sigma_{12}^2)$.} 
In that case, we can follow the first approach and plug in all the nuisance parameters. 

\begin{table}[p]
\begin{doublespacing}
\begin{center}
\scalebox{\shrinkageparameter}{
\begin{threeparttable}
{\scriptsize
\caption{Invertible Hypotheses in a Two-Factor Model}\label{table-2F-Hypotheses}
\begin{center}
\begin{tabular}{llc}
\hline\hline\vspace{-0.2cm}&&\\
Hypothesis&Further Reparameterization&Additional Assumption(s)\\
\hline\vspace{-0.2cm}\\
First FV&$r(\pi,\beta)=\sigma^2_1(\theta)$&$\lambda_{32}\neq 0$, $\lambda_{42}\neq 0$\\
Second FV&$r(\pi,\beta)=\beta$&-\\
FL $\lambda_{31}$&$r(\pi,\beta)=\lambda_{31}(\theta)$&$\rho_{42}\neq 0$, $\lambda_{32}\neq 0$\\
FL $\lambda_{41}$&$r(\pi,\beta)=\lambda_{41}(\theta)$&$\rho_{32}\neq 0$, $\lambda_{42}\neq 0$\\
FL $\lambda_{32}$&$r(\pi,\beta)=\lambda_{32}(\theta)$&$\rho_{41}\neq 0$, $\lambda_{32}\neq 0$\\
FL $\lambda_{42}$&$r(\pi,\beta)=\lambda_{42}(\theta)$&$\rho_{31}\neq 0$, $\lambda_{42}\neq 0$\\
First EV&$r(\pi,\beta)=\omega_1-\sigma^2_1(\theta)$&$\lambda_{32}\neq 0$, $\lambda_{42}\neq 0$\\
Second EV&$r(\pi,\beta)=\omega_2-\beta$&-\\
Third EV&$r(\pi,\beta)=\omega_3-\rho_{31}\lambda_{31}(\theta)-\rho_{32}\lambda_{32}(\theta)$&$\rho_{41}\rho_{32}-\rho_{42}\rho_{31}\neq 0$, $\lambda_{32}\neq 0$\\
Fourth EV&$r(\pi,\beta)=\omega_4-\rho_{41}\lambda_{41}(\theta)-\rho_{42}\lambda_{42}(\theta)$&$\rho_{31}\rho_{42}-\rho_{32}\rho_{41}\neq 0$, $\lambda_{42}\neq 0$\\
First StNR&$r(\pi,\beta)=(\omega_1-\sigma^2_1(\theta))\inv\sigma^2_1(\theta)$&$\lambda_{32}\neq 0$, $\lambda_{42}\neq 0$\\
Second StNR&$r(\pi,\beta)=(\omega_2-\beta)\inv \beta$&-\\
Third StNR&$r(\pi,\beta)=\frac{\rho_{31}\lambda_{31}(\theta)+\rho_{32}\lambda_{32}(\theta)}{\omega_3-\rho_{31}\lambda_{31}(\theta)-\rho_{32}\lambda_{32}(\theta)}$&$\rho_{41}\rho_{32}-\rho_{42}\rho_{31}\neq 0$, $\lambda_{32}\neq 0$\\
Fourth StNR&$r(\pi,\beta)=\frac{\rho_{41}\lambda_{41}(\theta)+\rho_{42}\lambda_{42}(\theta)}{\omega_4-\rho_{41}\lambda_{41}(\theta)-\rho_{42}\lambda_{42}(\theta)}$&$\rho_{31}\rho_{42}-\rho_{32}\rho_{41}\neq 0$, $\lambda_{42}\neq 0$\\
\hline
\end{tabular}
{\singlespace
\begin{tablenotes}
\item {\em Note:} The hypothesis is $H_0: r(\pi,\beta)=r_0$, where $r(\pi,\beta)$ is given in the second column, and $r_0$ is a hypothesized value. The formulas for $\sigma^2_1(\theta)$ and $\lambda_{jk}(\theta)$ for $j\in\{3,4\}$ and $k\in\{1,2\}$ are defined in (34)-(38). 
``FV'' stands for ``Factor Variance'', ``FL'' stands for ``Factor Loading'', ``EV'' stands for ``Error Variance'', and ``StNR'' stands for ``Signal to Noise Ratio''. 
The ``-'' indicates that no additional assumption is needed for $r(\pi,\beta)$ to be invertible for $\beta$. 
\end{tablenotes}
}
\end{center}
}
\end{threeparttable}
}
\end{center}
\end{doublespacing}
\end{table}

Table \ref{table-2F-Hypotheses} lists some hypotheses in the two-factor model that can be inverted for $\beta$, and any additional assumptions required. 
Table \ref{table-2F-Hypotheses} covers the variances of the factors, some factor loadings, error variances, and signal to noise ratios, which are commonly reported in current empirical practice. 

The hypotheses listed in Table \ref{table-2F-Hypotheses} do not take advantage of the possibility of switching the ordering of the observed variables. 
For example, if one wanted to test a factor loading on the fifth observed variable, one could assume $\lambda_{51}\neq 0$ and switch it for the fourth observed variable. 
Then, the additional assumption required for inverting the hypothesis would follow from the rows for $\lambda_{41}$ or $\lambda_{42}$ in Table \ref{table-2F-Hypotheses} with the appropriate switches made. 
Also, switching the first two variables is equivalent to interchanging the factors. 
Then, testing hypotheses on $\sigma^2_1$ (before the switch), for example, is equivalent to testing hypotheses on $\sigma^2_2$ (after the switch). 

\section{Simulations}

In this section, we use simulations to compare identification-robust hypothesis tests in factor models with one or two factors. 

\subsection{Identification-Robust Hypothesis Tests}

We compare various identification-robust hypothesis tests. 
We divide the tests into two types: ``original parameterization tests,'' which can be implemented without Reparameterizations \ref{reparameterization1} or \ref{reparameterization2}, and ``reparameterization tests,'' which require Reparameterizations \ref{reparameterization1} or \ref{reparameterization2} to be implemented. 
The original parameterization tests that we include are the projected \cite{AndersonRubin1949} test (AR-Proj) from \cite{StockWright2000}, the projected K test (K-Proj) from \cite{Kleibergen2005}, the two-step test (CZ) from \cite{ChaudhuriZivot2011}, the conditional linear combination test (CLC) from \cite{IAndrews2018}, and three two-step tests from \cite{DAndrews2017} denoted AR-AR, AR-LM, and AR-QLR. 
The reparameterization tests that we include are the plug-in \cite{AndersonRubin1949} test (AR-Plug) from \cite{StockWright2000}, the plug-in K test (K-Plug) and CLR test (CLR-Plug) from \cite{Kleibergen2005}, and the plug-in conditional likelihood ratio test (AM-Plug) from \cite{AndrewsMikusheva2016Functional}. 
\medskip

\textbf{Remarks on the Identification-Robust Tests.} 
\textbf{5.1.} \cite{IAndrews2018} focuses on identification-robust confidence sets. 
To convert this to a hypothesis test, we invert the robust confidence set defined in equation (12) in that paper, with $\gamma=0.05$ and $\alpha=0.05$. 
This ensures the nominal size of the CLC test is $0.05$ and the CLC test is comparable to the other identification-robust tests. 
This differs from the recommendation in \cite{IAndrews2018} of reporting two confidence sets, one identification-robust and one not, together with a coverage distortion size. 

\textbf{5.2.} The CZ test is computed with $\tau=0.045$ and $\zeta=0.005$ to ensure the nominal size is $0.05$ and the CZ test is comparable to the other identification-robust tests computed. 
For the same reason, the tests from \cite{DAndrews2017} are computed with $\alpha_1=0.005$ and $\alpha_2=0.045$. 

\textbf{5.3.} Several other identification-robust tests are omitted. 
This includes the versions of the AR, K, and CLR tests developed for empirical likelihood in \cite{GuggenbergerSmith2005, GuggenbergerSmith2008}, \cite{Otsu2006}, and \cite{GuggenbergerRamalhoSmith2012}. 
We expect these tests to be similar to the GMM versions of the AR, K, and CLR tests, and thus omit them. 
\cite{AndrewsMikusheva2016Geometric} propose a geometric test for curved null hypotheses in minimum distance models. 
Our hypotheses have unbounded curvature, and thus the test reduces to the AR-Proj test. 
\cite{AndrewsGuggenberger2019} propose identification and singularity-robust tests. 
Our moments have a non-singular variance matrix, and thus the tests should asymptotically reduce to the AR and CLR tests. 
\qed

\subsection{One-Factor Simulations}

In the one-factor model, we take the variance of the factor, $\sigma^2$, to be $1$. 
We take all the variances of the errors, $\phi_j$, to be $1$ for $j\in\{1,2,3\}$. 
Let $b\ge 0$. 
We consider two different specifications of the factor loadings. 
In the first specification, $\lambda_2=1$ and $\lambda_3=n^{-1/2}b$. 
In the second specification, $\lambda_2=\lambda_3=n^{-1/4}\sqrt{b}$. 
Note that the same value of $b$ should lead to the same ``strength'' of identification, measured in terms of $\sqrt{n}\rho_2\rho_3$. 
With these parameter values, $f_i$ and $\epsilon_i$ are simulated iid with a joint normal distribution. 
$W_i$ is calculated using (\ref{model}). 
In the simulations, we take $n=500$ and report results using $1000$ simulation draws. 
We test the hypothesis $H_0: \sigma^2=1.5$. 
When $b=0$, the parameter values are observationally equivalent to a vector of parameter values under the null, and thus the simulated rejection probabilities estimate a null rejection probability. 
When $b\neq 0$, the alternative hypothesis is true, and the simulated rejection probabilities estimate the power function. 

\begin{table}[p]
\begin{doublespacing}
\begin{center}
\scalebox{\shrinkageparameter}{
\begin{threeparttable}
{\scriptsize
\caption{Rejection Probabilities of Nominal 5\% Tests in a One-Factor Model}\label{table-1F-Power}
\begin{center}
\begin{tabular}{cccccccccccccc}
\hline\hline\vspace{-0.2cm}&&&&&&&&&&&&&\\
&\multicolumn{5}{c}{$(\lambda_2,\lambda_3)=(1,n^{-1/2}b)$}&&\multicolumn{5}{c}{$(\lambda_2,\lambda_3)=n^{-1/4}(\sqrt{b},\sqrt{b})$}&&\\
\cline{2-6}\cline{8-12}\vspace{-0.1cm}\\
{Test}\hspace{-0.08cm}&\hspace{-0.08cm}{$b\hspace{-0.09cm}=\hspace{-0.09cm}0$}\hspace{-0.08cm}&\hspace{-0.08cm}{$b\hspace{-0.09cm}=\hspace{-0.09cm}1$}\hspace{-0.08cm}&\hspace{-0.08cm}{$b\hspace{-0.09cm}=\hspace{-0.09cm}2$}\hspace{-0.08cm}&\hspace{-0.08cm}{$b\hspace{-0.09cm}=\hspace{-0.09cm}5$}\hspace{-0.08cm}&\hspace{-0.08cm}{$b\hspace{-0.09cm}=\hspace{-0.09cm}10$}\hspace{-0.08cm}&\hspace{-0.08cm}\hspace{-0.08cm}&\hspace{-0.08cm} {$b\hspace{-0.09cm}=\hspace{-0.09cm}0$}\hspace{-0.08cm}&\hspace{-0.08cm}{$b\hspace{-0.09cm}=\hspace{-0.09cm}1$}\hspace{-0.08cm}&\hspace{-0.08cm}{$b\hspace{-0.09cm}=\hspace{-0.09cm}2$}\hspace{-0.08cm}&\hspace{-0.08cm}{$b\hspace{-0.09cm}=\hspace{-0.09cm}5$}\hspace{-0.08cm}&\hspace{-0.08cm}{$b\hspace{-0.09cm}=\hspace{-0.09cm}10$}\hspace{-0.08cm}&\hspace{-0.08cm}\hspace{-0.08cm}&\hspace{-0.08cm}{Time}\\
\hline\vspace{-0.2cm}\\
\multicolumn{14}{c}{Original Parameterization Tests}\\
\hline\vspace{-0.2cm}\\
AR-Proj&0.1&0.1&0.1&1.1&9.4&&0.1&0.1&0.2&1.7&11.0&&0.01\\
CZ&2.4&3.3&4.1&14.9&50.8&&2.7&3.8&5.6&19.9&55.7&&0.06\\
CLC&4.0&4.4&6.1&18.8&54.6&&4.1&4.8&7.4&23.3&58.9&&0.04\\
AR-AR&2.1&2.6&3.8&13.9&48.9&&2.8&3.2&4.8&18.6&53.6&&0.04\\
AR-LM&0&0&0&0&0.4&&0&0&0&0&1.1&&0.01\\
\hline\vspace{-0.2cm}\\
\multicolumn{14}{c}{Reparameterization Tests}\\
\hline\vspace{-0.2cm}\\
AR-Plug&5.7&6.1&8.0&23.2&62.2&&5.9&6.2&9.7&28.1&66.4&&0.01\\
AM-Plug&11.0&12.8&15.0&27.7&63.9&&10.9&9.8&14.2&29.0&67.5&&192\\
\hline
\vspace{-0.2cm}\\
Ave. Len. &3.8&3.7&3.4&1.8&0.8&&3.8&3.5&3.0&1.5&0.7&&\\
\hline
\end{tabular}
{\singlespace
\begin{tablenotes}
\item {\em Note:} The parameters in the data generating process are $\sigma^2=\phi_1=\phi_2=\phi_3=1$ and $n=500$. 
This implies that $\sqrt{n}\rho_2\rho_3=b$ for all specifications. 
The entries in the table denote the rejection probabilities for testing $H_0: \sigma^2=1.5$, reported in percentages out of 1000 simulations. 
``Ave. Len.'' denotes the average length of the confidence interval formed by inverting the AR-Plug test. 
The entries in the ``Time'' column report the average time to compute each test in seconds per simulation. 
\end{tablenotes}
}
\end{center}
}
\end{threeparttable}
}
\end{center}
\end{doublespacing}
\end{table}

Table \ref{table-1F-Power} reports the simulated rejection probabilities for the identification-robust tests in the one-factor model with $b\in\{0,1,2,5,10\}$. \medskip

\textbf{Remarks on Table \ref{table-1F-Power}.} \textbf{5.4.} The CLR-Plug and K-Plug tests are omitted because they reduce to the AR-Plug test when the number of moments is equal to the number of parameters. 
Similarly, the K-Proj test is omitted because it reduces to the AR-Proj test. 

\textbf{5.5.} The columns with $b=0$ correspond to null rejection probabilities. 
We see that the AR-Proj and AR-LM tests are extremely conservative, the CZ and AR-AR tests are moderately conservative, and the CLC test is slightly conservative. 
Conversely, the AR-Plug test has slight over-rejection under the null, and the AM-Plug test has significant finite-sample over-rejection.\footnote{In unreported simulations, we increased the sample size and found that the null rejection probability of the AM-Plug test approaches 5\%. This suggests the over-rejection of the AM-Plug test is a finite-sample result.} 

\textbf{5.6.} The columns with $b>0$ show the power function of the tests. 
As the strength of identification increases, the power of the tests increases. 
The AM-Plug and AR-Plug tests have the highest power, followed by the CLC, CZ, and AR-AR tests. 
This ranking follows the ranking based on null rejection probability. 
Overall, the AR-Plug and CLC tests provide a good trade-off between size and power. 

\textbf{5.7.} Table \ref{table-1F-Power} reports the average length of the confidence interval formed by inverting the AR-Plug test. 
This is to provide some intuition on the scale of $b$. 
Because it is a local parameter, $b$ does not have easily interpretable units. 

\textbf{5.8.} We also note that the AM-Plug test is much more computationally expensive than the other tests. 
\qed \medskip

We also investigate the problem of estimating the number of factors in these specifications. 
We consider estimators that come from the model selection literature for GMM following \cite{Andrews1999}. 
With only three observed variables, we compare a zero-factor model and a one-factor model and choose the one that minimizes AIC or BIC. 
We also report the probability of rejecting the specification of the zero-factor model using a GMM J-test. 
Table \ref{table-1F-Number} reports these estimates for $b\in\{0,1/3,2/3,1,2\}$. \medskip

\begin{table}[p]
\begin{doublespacing}
\begin{center}
\scalebox{\shrinkageparameter}{
\begin{threeparttable}
{\scriptsize
\caption{Estimates of the Number of Factors in a One-Factor Model}\label{table-1F-Number}
\begin{center}
\begin{tabular}{cccccccccccc}
\hline\hline\vspace{-0.2cm}&&&&&&&&&&&\\
&\multicolumn{5}{c}{$(\lambda_2,\lambda_3)=(1,n^{-1/2}b)$}&&\multicolumn{5}{c}{$(\lambda_2,\lambda_3)=n^{-1/4}(\sqrt{b},\sqrt{b})$}\\
\cline{2-6}\cline{8-12}\vspace{-0.2cm}\\
&{$b=0$}&{$b=1/3$}&{$b=2/3$}&{$b=1$}&{$b=2$}&& {$b=0$}&{$b=1/3$}&{$b=2/3$}&{$b=1$}&{$b=2$}\\
\hline\vspace{-0.2cm}\\
AIC&100&100&100&100&100&&12.3&76.1&95.7&99.4&100\\
BIC&100&100&100&100&100&&0&11.7&45.6&75.2&99.3\\
J-Test&100&100&100&100&100&&5.4&64.0&91.2&98.6&100\\
\hline
\end{tabular}
{\singlespace
\begin{tablenotes}
\item {\em Note:} The parameters in the data generating process are $\sigma^2=\phi_1=\phi_2=\phi_3=1$ and $n=500$. 
This implies that $\sqrt{n}\rho_2\rho_3=b$ for all specifications. 
The entries for rows AIC and BIC denote the percentage of simulations out of 1000 for which one factor was estimated. 
One minus the entry gives the percentage of simulations for which zero factors were estimated.
The entry for the J-Test row denotes the percentage of simulations out of 1000 for which the GMM J-Test for the specification of a zero-factor model rejects. 
\end{tablenotes}
}
\end{center}
}
\end{threeparttable}
}
\end{center}
\end{doublespacing}
\end{table}

\textbf{Remarks on Table \ref{table-1F-Number}.} \textbf{5.9.} When $\lambda_2=1$, AIC and BIC both estimate one factor always. 
The J-test for the specification of the zero-factor model also rejects always. 
Thus, the number of factors is identified under weaker conditions than the model parameters. 
Weak identification may be a problem even if the number of factors is estimated correctly. 

\textbf{5.10.} When both $\lambda_2$ and $\lambda_3$ are close to zero, the AIC and BIC estimates do not always detect the factors. 
Thus, weak identification can lead to the number of factors being under-estimated. 

\textbf{5.11.} In the data generating process for Table \ref{table-1F-Number}, it is not possible for the number of factors to be over-estimated. 
This is because no over-parameterized models are considered. 
This may preference the results in favor of AIC over BIC. 
However, Table \ref{table-1F-Number} should not be understood to be comparing AIC and BIC. 
Instead, Table \ref{table-1F-Number} is just to investigate when AIC and BIC correctly estimate the number of factors in the presence of weak identification. 
\qed

\subsection{Two-Factor Simulations}
\label{two-factor-simulations}

In the two-factor model, we take the covariance matrix of the factors to be $\Sigma=I_2$. 
We take all the variances of the errors, $\phi_j$, to be $1$ for $j\in\{1,2,3,4,5\}$. 
Let $b_1, b_2\ge 0$. 
We consider three different specifications of the factor loadings. 
In the first specification, $\lambda_{31}=\lambda_{41}=\lambda_{52}=1$, $\lambda_{51}=0$, and $(\lambda_{32},\lambda_{42})=n^{-1/2}(b_1,b_2)$. 
This specification is weakly identified because the second factor only has two nonzero factor loadings in the limit. 
In the second specification, $\lambda_{31}=\lambda_{41}=\lambda_{51}=\lambda_{52}=1$ and $(\lambda_{32},\lambda_{42})=(1-n^{-1/2}b_2,1-n^{-1/2}b_1)$. 
This specification is weakly identified because the two factors cannot be separately identified because $(\lambda_{31},\lambda_{41},\lambda_{51})$ is collinear with $(\lambda_{32},\lambda_{42},\lambda_{52})$ in the limit. 
In the third specification, $\lambda_{31}=\lambda_{41}=1$, $\lambda_{51}=0$, and $(\lambda_{32},\lambda_{42},\lambda_{52})=n^{-1/4}(\sqrt{b_1},b_1^{-1/2}b_2,\sqrt{b_1})$. 
(When $b_1=0$, we take $b_1^{-1/2}b_2=0$.) 
This specification is weakly identified because the second factor only has one nonzero factor loading in the limit. 
Also note that the convergence rate is slower relative to the first specification. 
In all specifications, the same value of $(b_1,b_2)$ should lead to the same strength of identification, measured in terms of $\sqrt{n}(s_1(\pi),s_2(\pi))$. 
With these parameter values, $f_i$ and $\epsilon_i$ are simulated iid with a joint normal distribution. 
$W_i$ is calculated using (\ref{model}). 
In the simulations, we take $n=500$ and report results using $1000$ simulation draws. 
We test the hypothesis $H_0: \sigma^2_2=1.5$. 
When $(b_1,b_2)=(0,0)$, the parameter values are observationally equivalent to a vector of parameter values under the null, and thus the simulated rejection probabilities estimate a null rejection probability. 
When $(b_1,b_2)\neq (0,0)$, the alternative hypothesis is true, and the simulated rejection probabilities estimate the power function. 

\begin{table}[p]
\begin{doublespacing}
\begin{center}
\vspace{-0.4cm}
\rotatebox{270}{
\begin{threeparttable}
{\scriptsize
\caption{Rejection Probabilities of Nominal 5\% Tests in a Two-Factor Model}\label{table-2F-Power}
\begin{center}
\begin{tabular}{cccccccccccccccccccc}
\hline\hline&&&&&&&&&&&&&&&&&&&\\
&\multicolumn{5}{c}{$\lambda_{32}=\frac{b_1}{\sqrt{n}}$, $\lambda_{42}=0$, $\lambda_{51}=0$, $\lambda_{52}=1$}&\hspace{-0.11cm}&\multicolumn{5}{c}{$\lambda_{32}=1$, $\lambda_{42}=1-\frac{b_1}{\sqrt{n}}$, $\lambda_{51}=1$, $\lambda_{52}=1$}&\hspace{-0.11cm}&\multicolumn{5}{c}{$(\lambda_{32},\lambda_{52})=\frac{(\sqrt{b_1},\sqrt{b_1})}{n^{1/4}}$, $\lambda_{42}=0$, $\lambda_{51}=0$}&\\
\cline{2-6}\cline{8-12}\cline{14-18}\\
{Test}\hspace{-0.18cm}&\hspace{-0.18cm}{$b_1=0$}&{$b_1=2$}&{$b_1=5$}&{$b_1=10$}&{$b_1=20$}&\hspace{-0.11cm}&{$b_1=0$}&{$b_1=2$}&{$b_1=5$}&{$b_1=10$}&{$b_1=20$}&\hspace{-0.11cm}&{$b_1=0$}&{$b_1=2$}&{$b_1=5$}&{$b_1=10$}&{$b_1=20$}\hspace{-0.18cm}&\hspace{-0.11cm}&\hspace{-0.18cm}{Time}\hspace{-0.18cm}\\
\hline\\
\multicolumn{20}{c}{Original Parameterization Tests}\\
\hline\\
AR-Proj\hspace{-0.18cm}&\hspace{-0.18cm}0&0&0.1&0.4&2.6&&0&0&0&0&0.9&&0&0&0.1&0.4&2.7&&\hspace{-0.18cm}0.05\\
K-Proj\hspace{-0.18cm}&\hspace{-0.18cm}0&0&0.1&0.2&2.2&&0&0&0&0&0.8&&0&0&0.1&0.2&2.5&&\hspace{-0.18cm}0.25\\
CZ\hspace{-0.18cm}&\hspace{-0.18cm}0&0&0.1&3.9&59.4&&0&0&0&0.3&42.9&&0&0.1&1.4&14.7&60.6&&\hspace{-0.18cm}0.16\\
CLC\hspace{-0.18cm}&\hspace{-0.18cm}0&0&0&0.1&28.0&&0&0&0&0&22.4&&0&0&0&1.2&30.4&&\hspace{-0.18cm}0.36\\
AR-AR\hspace{-0.18cm}&\hspace{-0.18cm}0.5&1.0&2.2&11.4&54.2&&0.5&0.7&1.5&5.2&30.6&&0.7&1.1&2.7&14.0&54.8&&\hspace{-0.18cm}0.09\\
AR-LM\hspace{-0.18cm}&\hspace{-0.18cm}0&0&0&0&0.8&&0&0&0&0&0&&0&0&0&0&0.8&&\hspace{-0.18cm}0.06\\
AR-QLR\hspace{-0.18cm}&\hspace{-0.18cm}0&0&0&0.2&2.1&&0&0&0&0&0.7&&0&0&0.1&0.3&2.2&&\hspace{-0.18cm}0.06\\
\hline\\
\multicolumn{20}{c}{Reparameterization Tests}\\
\hline\\
AR-Plug\hspace{-0.18cm}&\hspace{-0.18cm}6.0&7.0&14.0&38.7&77.3&&4.9&5.8&8.8&19.5&57.5&&5.9&8.0&16.7&43.6&77.5&&\hspace{-0.18cm}0.04\\
K-Plug\hspace{-0.18cm}&\hspace{-0.18cm}4.1&6.1&15.1&47.3&83.9&&4.7&5.1&8.3&25.6&69.3&&3.7&6.8&21.7&52.4&84.3&&\hspace{-0.18cm}0.05\\
CLR-Plug\hspace{-0.18cm}&\hspace{-0.18cm}5.9&6.5&14.6&47.4&84.1&&5.0&5.9&9.1&25.4&69.5&&5.7&7.3&19.4&53.3&84.2&&\hspace{-0.18cm}0.91\\
\hline
\\
Ave. Len.\hspace{-0.18cm}&\hspace{-0.18cm}9.3&8.6&4.7&1.1&0.7&&9.4&9.4&7.9&3.4&0.8&&9.5&7.9&3.4&1.0&0.6&&\\
\hline\vspace{-0.1cm}
\end{tabular}
{\singlespace
\begin{tablenotes}
\item {\em Note:} The parameters in the data generating process are $\sigma^2_1=\sigma^2_2=\phi_1=\phi_2=\phi_3=\phi_4=\phi_5=\lambda_{31}=\lambda_{41}=1$, $\sigma_{12}=0$, $b_2=0$, and $n=500$. 
This implies that $\sqrt{n}(s_1(\pi),s_2(\pi))=(b_1,b_2)+o(1)$ for all specifications. 
The entries in the table denote the rejection probabilities for testing $H_0: \sigma^2_2=1.5$, reported in percentages out of 1000 simulations. 
``Ave. Len.'' denotes the average length of the confidence interval formed by inverting the AR-Plug test. 
The entries in the ``Time'' column report the average time to compute each test in seconds per simulation. 
\end{tablenotes}
}
\end{center}
}
\end{threeparttable}
}
\end{center}
\end{doublespacing}
\end{table}

Table \ref{table-2F-Power} reports the simulated rejection probabilities for the identification-robust tests in the two-factor model with $b_1\in\{0,2,5,10,20\}$ and $b_2=0$. \medskip

\textbf{Remarks on Table \ref{table-2F-Power}.} \textbf{5.12.} The AM-Plug test is omitted because of computational cost. 
A projected version of the CLR test is omitted because (a) we expect it to be very conservative like the AR-Proj and K-Proj tests, and (b) it is computationally difficult to project the CLR test because one must recalculate the critical value at each point. 

\textbf{5.13.} The columns with $(b_1,b_2)=(0,0)$ correspond to null rejection probabilities. 
We see that the reparameterization tests have reasonable rejection probabilities under the null. 
On the contrary, all the original parameterization tests are extremely conservative. 
This is expected for the projected versions of the full-vector tests. 
This is somewhat surprising for the CZ, CLC, AR-AR, AR-LM, and AR-QLR tests, which include modifications to correct for the conservativeness of the projections. 
The CZ, CLC, AR-LM, and AR-QLR tests are based on an efficient LM statistic, which is designed to be efficient under strong identification. 
In addition, the CZ, AR-AR, AR-LM, and AR-QLR tests restrict the projection to a first-stage confidence set, which should not lead to conservativeness under strong identification. 
Under weak identification, however, the validity of the tests still rely on projection. 
The CLC test is based on a linear combination of the AR statistic and the efficient LM statistic, which is shown to be admissible for a full-vector hypothesis in \cite{Andrews2016}. 
Still, the test must be projected over all the nuisance parameters. 
While these modifications work well under strong identification, they are unable to overcome the conservativeness under weak identification. 

\textbf{5.14.} The columns with $b_1>0$ show the power function of the tests. 
As the strength of identification increases, measured using $b_1$, the power of the tests increases. 
The AR-Plug, K-Plug, and CLR-Plug tests all have very similar power. 
The AR-Plug test has slightly lower power, while the CLR-Plug test takes somewhat more time to compute. 
Overall, the reparameterization tests are recommended, with the K-Plug test performing particularly well. 

\textbf{5.15.} Table \ref{table-2F-Power} reports the average length of the confidence interval formed by inverting the AR-Plug test. 
This is to provide some intuition on the scale of $b_1$ and $b_2$. 
Because they are local parameters, $b_1$ and $b_2$ do not have easily interpretable units.
\qed \medskip

We also investigate the problem of estimating the number of factors in these specifications. 
We consider estimators that come from the model selection literature for GMM following \cite{Andrews1999}. 
With five observed variables, we compare a one-factor model and a two-factor model and choose the one that minimizes AIC or BIC. 
We also report the probability of rejecting the specification of the one-factor model using a GMM J-test. 
Table \ref{table-2F-Number} reports these estimates for $b_1\in\{0,1,1.5,2,5\}$ and $b_2=0$. \medskip

\begin{table}[p]
\begin{doublespacing}
\begin{center}
\scalebox{\shrinkageparameter}{
\begin{threeparttable}
{\scriptsize
\caption{Estimates of the Number of Factors in a Two-Factor Model}\label{table-2F-Number}
\begin{center}
\begin{tabular}{cccccc}
\hline\hline\vspace{-0.2cm}&&&&&\\
&\multicolumn{5}{c}{$(\lambda_{32},\lambda_{52})=n^{-1/4}(\sqrt{b_1},\sqrt{b_1})$, $\lambda_{42}=0$, $\lambda_{51}=0$}\\
\cline{2-6}\vspace{-0.2cm}\\
&{\hspace{5mm}$b_1=0$\hspace{5mm}}&{\hspace{5mm}$b_1=1$\hspace{5mm}}&{\hspace{5mm}$b_1=1.5$\hspace{5mm}}&{\hspace{5mm}$b_1=2$\hspace{5mm}}&{\hspace{5mm}$b_1=5$\hspace{5mm}}\\
\hline\vspace{-0.2cm}\\
AIC&17.4&96.6&99.6&99.9&100\\
BIC&10.9&47.7&72.2&87.3&99.9\\
J-Test&15.1&92.0&98.8&99.9&100\\
\hline
\end{tabular}
{\singlespace
\begin{tablenotes}
\item {\em Note:} The parameters in the data generating process are $\sigma^2_1=\sigma^2_2=\phi_1=\phi_2=\phi_3=\phi_4=\phi_5=\lambda_{31}=\lambda_{41}=1$, $\sigma_{12}=\lambda_{51}=0$, $b_2=0$, and $n=500$. 
This implies that $\sqrt{n}(s_1(\pi),s_2(\pi))=(b_1,b_2)$ for all specifications. 
The entries for rows AIC and BIC denote the percentage of simulations out of 1000 for which two factors were estimated. 
One minus the entry gives the percentage of simulations for which one factor was estimated.
The entry for the J-Test row denotes the percentage of simulations out of 1000 for which the GMM J-Test for the specification of a one-factor model rejects. 
\end{tablenotes}
}
\end{center}
}
\end{threeparttable}
}
\end{center}
\end{doublespacing}
\end{table}

\textbf{Remarks on Table \ref{table-2F-Number}.} \textbf{5.16.} The estimates for the first two specifications are omitted. 
This is because, in those specifications, AIC and BIC both estimate two factors always. 
The J-test for the specification of the one-factor model also rejects always. 
As with the one-factor model, the number of factors is identified under weaker conditions than the model parameters. 
Weak identification may be a problem even if the number of factors is estimated correctly. 

\textbf{5.17.} In the third specification, when $\lambda_{32}$ and $\lambda_{52}$ both converge to zero at the $n^{-1/4}$ rate, the AIC and BIC estimates do not always detect the second factor. 
Thus, weak identification can lead to the number of factors being under-estimated. 
\qed

\section{Empirical Application}

\cite{Attanasio2014} report improved cognition and language development in the treated children following a randomized intervention in Colombia. 
The intervention consisted of 18 months of weekly home visits designed to ``improve the quality of maternal-child interactions by teaching the mothers to participate in developmentally appropriate learning activities.''\footnote{\cite{Attanasio2014}, pg. 3.} 
The dataset includes a variety of variables measuring time spent with the child and material investments in the child. 
\cite{Attanasio2020AER} specify a factor model for these variables and give a structural interpretation to the common factor, labeling it ``parental investments.'' 
In the literature on childhood development, researchers are interested in the distribution of parental investment factors over time and across populations.\footnote{Parental investment factors can be used as inputs into the production function for human capital; see \cite{CunhaHeckman2008}, \cite{CunhaHeckmanSchennach2010}, \cite{Attanasio2020AER, Attanasio2020ReStud}, and \cite{AgostinelliWiswall2021}.} 

\cite{Attanasio2020AER} consider a factor model for the variables measuring time and material investments. 
Table C.1 in \cite{Attanasio2020AER} reports a variety of estimates of the number of factors. 
The estimates range from one to four factors. 
Ultimately, \cite{Attanasio2020AER} specify a model with one factor for the treatment group and one factor for the control group. 
We consider a factor model for only the variables measuring material investments, which are the number of different kinds of toys. 
The variables are: (1) the number of total types of play materials, (2) the number of coloring/drawing books, (3) the number of toys to learn movement, (4) the number of shop-bought toys, and (5) the number of toys to learn shapes.\footnote{Details on the collection of these variables can be found in Section B.3 in \cite{Attanasio2020AER}. In particular, ``Play materials include toys that make/play music; toys/objects meant for stacking, constructing or building; things for drawing, writing, colouring, and painting; toys for moving around; toys to play pretend games; picture and drawing books for children; and toys for learning shapes and colours.'' (\cite{Attanasio2020AER}, Section B.3, pg. 10).} 
We find evidence of two factors for both the treatment and control groups. 
The AIC and BIC both estimate two factors, and the J-tests for the specification of a one-factor model reject with p-values less than $10^{-8}$ for both groups. 
 
We consider two-factor models for the treatment and control groups allowing for weak identification. 
As in Section \ref{Section2}, we assume the factor loadings for the first two variables is the $2\times 2$ identity matrix. 
This requires the first variable to not depend on the second factor and the second variable to not depend on the first factor. 
One can think of the two factors as indexing different characteristics of households. 
The first (second) factor indicates households that have a higher value of the first (second) variable, without changing the second (first) variable. 
We also assume the third and fourth variables have nonzero factor loadings on both factors. 
This can be justified by arguing that the third and fourth variables have a common component with the first and second variables. 
For example, the assumption that the third variable has a nonzero factor loading on the first factor can be justified by arguing that the first variable (the number of total types of play materials) and the third variable (the number of toys to learn movement) contribute to a common component of parental investment. 
Put another way, giving a child an additional type of play material and giving a child a toy to learn movement are both contributions to a common component of investment in the skills of the child. 
As discussed in Section \ref{Section4}, these assumptions are insufficient for identification. 
The model may be weakly identified even with nonzero factor loadings if, for example, there is collinearity between the factor loadings for the first and second factor. 

\cite{Attanasio2020AER} report point estimates and standard errors for factor model parameters, including the factor loadings and the variances of the errors. 
Let $\sigma^2_{c,1}$ and $\sigma^2_{c,2}$ denote the variances of the factors for the control group, and let $\sigma^2_{t,1}$ and $\sigma^2_{t,2}$ denote the variances of the factors for the treatment group. 
The variances of the factors are interesting because they represent the heterogeneity of parental investments across households. 
Table \ref{EmpiricalResults2} reports point estimates and confidence intervals (CIs) for the variances of the factors. 
Table \ref{EmpiricalResults2} includes a standard CI calculated using a t-statistic, an original-parameterization identification-robust CI formed by inverting the AR-AR test, and a reparameterization identification-robust CI formed by inverting the CLR-Plug test. 
For $\sigma^2_{c,1}$ and $\sigma^2_{t,1}$, we use the further reparameterization in Table \ref{table-2F-Hypotheses} to plug in all the nuisance parameters. 
The AR-AR test is chosen to represent the original-parameterization identification-robust CIs because it is the least conservative in the simulations with two factors in Section \ref{two-factor-simulations}. 
Table 8, in the Supplemental Materials, reports point estimates and identification-robust confidence intervals for the factor loadings. 

\begin{table}[p]
\begin{doublespacing}
\begin{center}
\scalebox{\shrinkageparameter}{
\begin{threeparttable}
{\scriptsize
\caption{Parental Investment Factors}\label{EmpiricalResults2}
\begin{tabular}{lccccc}
\hline\hline\vspace{-0.2cm}\\
&\multicolumn{2}{c}{Control}&&\multicolumn{2}{c}{Treatment}\\
\cline{2-3}\cline{5-6}\vspace{-0.2cm}\\
&{$\sigma^2_{c,1}$}&{$\sigma^2_{c,2}$}&&{$\sigma^2_{t,1}$}&{$\sigma^2_{t,2}$}\\
\hline\vspace{-0.2cm}\\
Point Estimate&0.99&0.28&&1.01&0.08\\
Standard CI&[0.85,1.14]&[0.17,0.39]&&[0.84,1.18]&[0.04,0.12]\\
Orig.-Param. Id.-Robust CI&[0.79,1.71]&[0.09,0.41]&&[0.92,10]&[0.001,0.12]\\
Reparam. Id.-Robust CI&[0.90,1.36]&[0.18,0.34]&&[0.98,10]&[0.02,0.08]\\
\hline
\end{tabular}
{\singlespace
\begin{tablenotes}
\item {\em Note:} The point estimate minimizes the GMM objective function. 
The standard CI is calculated using a t-statistic. 
The original parameterization identification-robust CI is calculated by inverting the AR-AR test. 
The reparameterization identification-robust CI is calculated by inverting the CLR-Plug test. 
\end{tablenotes}
}
}
\end{threeparttable}
}
\end{center}
\end{doublespacing}
\end{table}

\textbf{Remarks.} \textbf{6.1.} The standard CI is only valid under strong identification. 
The identification-robust CIs are valid under both strong and weak identification. 
Here, some of the identification-robust CIs are substantially wider than the standard CI, indicating that weak identification is relevant. 
Still, the identification-robust confidence intervals are not necessarily wider than the standard CI. 
They are not nested. 

\textbf{6.2.} The original-parameterization identification-robust CIs contain the reparameterization identification-robust CIs. 
This is consistent with the simulation results, which show that the CLR-Plug test is less conservative and more powerful than the AR-AR test. 

\textbf{6.3.} The identification-robust CIs can be exceptionally long, especially the CI for $\sigma^2_{t,1}$. 
For $\sigma^2_{t,1}$, the upper bound of both identification-robust CIs is 10. 
This upper bound is an arbitrary truncation of the parameter values considered. 
If larger parameter values were considered, the upper bound would probably be larger and possibly unbounded. 
This is a common feature of identification-robust CIs under weak identification. 
\cite{Cox_weak_id_w_bounds} shows how to use additional inequalities to make identification-robust CIs more informative in this case. 
In factor models, useful inequalities come from nonnegativity of the variances of the idiosyncratic errors. 
\cite{Cox_weak_id_w_bounds} shows that using these inequalities leads to identification-robust CIs that are reasonable in length. 
\qed  

\section{Conclusion}

This paper describes how to reparameterize low-dimensional factor models with one or two factors to fit the weak identification theory developed for GMM models. 
The reparameterizations are useful for identification-robust hypothesis testing. 
Simulations and an empirical application show the benefit of using the reparameterizations in order to use identification-robust hypothesis tests that require strongly identified nuisance parameters. 

\bibliography{references}

\end{document}